\documentclass[11pt,a4paper]{article}
\usepackage{ifthen} % for conditional statements
\newboolean{pdflatex}
\setboolean{pdflatex}{true} % False for eps figures 

\newboolean{articletitles}
\setboolean{articletitles}{true} % False removes titles in references

\newboolean{uprightparticles}
\setboolean{uprightparticles}{false} %True for upright particle symbols

\newboolean{inbibliography}
\setboolean{inbibliography}{false} %True once you enter the bibliography

\usepackage{microtype}
\usepackage{lineno}  % for line numbering during review
\usepackage{xspace} % To avoid problems with missing or double spaces after
\usepackage{caption} %these three command get the figure and table captions automatically small

\usepackage{graphicx}  % to include figures (can also use other packages)
\usepackage{color}
\usepackage{colortbl}
\graphicspath{{./figs/}} % Make Latex search fig subdir for figures
\usepackage{pdflscape}

\usepackage{todonotes}

\usepackage[abbreviations = true]{siunitx} %proper units
\usepackage{amsmath} % Adds a large collection of math symbols
\usepackage{amssymb}
\usepackage{amsfonts}
\usepackage{upgreek} % Adds in support for greek letters in roman typeset

\usepackage{jinstpub}
\usepackage{xspace} 
\usepackage{upgreek}

\def\MagUp {\mbox{\em Mag\kern -0.05em Up}\xspace}

\ifthenelse{\boolean{uprightparticles}}%
{

 \def\PDelta      {\ensuremath{\Delta}\xspace}                 
 \def\PXi      {\ensuremath{\Xi}\xspace}                 
 \def\PLambda      {\ensuremath{\Lambda}\xspace}                 
 \def\PSigma      {\ensuremath{\Sigma}\xspace}                 
 \def\POmega      {\ensuremath{\Omega}\xspace}                 
 \def\PUpsilon      {\ensuremath{\Upsilon}\xspace}                 
 
 \def\PB      {\ensuremath{\mathrm{B}}\xspace}                 
                  
 \def\PD      {\ensuremath{\mathrm{D}}\xspace}

 \def\PK      {\ensuremath{\mathrm{K}}\xspace}

 \def\Pe      {\ensuremath{\mathrm{e}}\xspace}

 \def\Pi      {\ensuremath{\mathrm{i}}\xspace}

}
{

 \mathchardef\PDelta="7101
 \mathchardef\PXi="7104
 \mathchardef\PLambda="7103
 \mathchardef\PSigma="7106
 \mathchardef\POmega="710A
 \mathchardef\PUpsilon="7107
                  
 \def\PB      {\ensuremath{B}\xspace}                 
                  
 \def\PD      {\ensuremath{D}\xspace}

 \def\PK      {\ensuremath{K}\xspace}

 \def\Pe      {\ensuremath{e}\xspace}

 \def\Pi      {\ensuremath{i}\xspace}

}

\makeatletter
\ifcase \@ptsize \relax% 10pt
  \newcommand{\miniscule}{\@setfontsize\miniscule{4}{5}}% \tiny: 5/6
\or% 11pt
  \newcommand{\miniscule}{\@setfontsize\miniscule{5}{6}}% \tiny: 6/7
\or% 12pt
  \newcommand{\miniscule}{\@setfontsize\miniscule{5}{6}}% \tiny: 6/7
\fi
\makeatother

\DeclareRobustCommand{\optbar}[1]{\shortstack{{\miniscule (\rule[.5ex]{1.25em}{.18mm})}
  \\ [-.7ex] $#1$}}

\def\electron   {{\ensuremath{\Pe}}\xspace}
\def\en         {{\ensuremath{\Pe^-}}\xspace}   % electron negative (\em is taken)

  \def\Kbar    {{\kern 0.2em\overline{\kern -0.2em \PK}{}}\xspace}

\def\KorKbar    {\kern 0.18em\optbar{\kern -0.18em K}{}\xspace}

  \def\Dbar    {{\kern 0.2em\overline{\kern -0.2em \PD}{}}\xspace}

\def\DorDbar    {\kern 0.18em\optbar{\kern -0.18em D}{}\xspace}

\def\Bbar    {{\ensuremath{\kern 0.18em\overline{\kern -0.18em \PB}{}}}\xspace}

\def\BorBbar    {\kern 0.18em\optbar{\kern -0.18em B}{}\xspace}

  \def\Y#1S{\ensuremath{\PUpsilon{(#1S)}}\xspace}% no space before {...}!

\def\Lbar        {{\ensuremath{\kern 0.1em\overline{\kern -0.1em\PLambda}}}\xspace}
\def\LorLbar    {\kern 0.18em\optbar{\kern -0.18em \PLambda}{}\xspace}

\def\AT#1     {\ensuremath{A_{\mathrm{T}}^{#1}}\xspace}           % 2

\def\C#1      {\ensuremath{\mathcal{C}_{#1}}\xspace}                       % 9
\def\Cp#1     {\ensuremath{\mathcal{C}_{#1}^{'}}\xspace}                    % 7
\def\Ceff#1   {\ensuremath{\mathcal{C}_{#1}^{\mathrm{(eff)}}}\xspace}        % 9  
\def\Cpeff#1  {\ensuremath{\mathcal{C}_{#1}^{'\mathrm{(eff)}}}\xspace}       % 7
\def\Ope#1    {\ensuremath{\mathcal{O}_{#1}}\xspace}                       % 2
\def\Opep#1   {\ensuremath{\mathcal{O}_{#1}^{'}}\xspace}                    % 7

\newcommand{\tev}{\ifthenelse{\boolean{inbibliography}}{\ensuremath{~T\kern -0.05em eV}\xspace}{\ensuremath{\mathrm{\,Te\kern -0.1em V}}}\xspace}
\newcommand{\gev}{\ensuremath{\mathrm{\,Ge\kern -0.1em V}}\xspace}
\newcommand{\mev}{\ensuremath{\mathrm{\,Me\kern -0.1em V}}\xspace}
\newcommand{\kev}{\ensuremath{\mathrm{\,ke\kern -0.1em V}}\xspace}
\newcommand{\ev}{\ensuremath{\mathrm{\,e\kern -0.1em V}}\xspace}
\newcommand{\gevc}{\ensuremath{{\mathrm{\,Ge\kern -0.1em V\!/}c}}\xspace}
\newcommand{\mevc}{\ensuremath{{\mathrm{\,Me\kern -0.1em V\!/}c}}\xspace}
\newcommand{\gevcc}{\ensuremath{{\mathrm{\,Ge\kern -0.1em V\!/}c^2}}\xspace}
\newcommand{\gevgevcccc}{\ensuremath{{\mathrm{\,Ge\kern -0.1em V^2\!/}c^4}}\xspace}
\newcommand{\mevcc}{\ensuremath{{\mathrm{\,Me\kern -0.1em V\!/}c^2}}\xspace}

\def\cm   {\ensuremath{\mathrm{ \,cm}}\xspace}

\def\mum  {\ensuremath{{\,\upmu\mathrm{m}}}\xspace}

\def\nm   {\ensuremath{\mathrm{ \,nm}}\xspace}

\def\invfb   {\ensuremath{\mbox{\,fb}^{-1}}\xspace}

\def\sec  {\ensuremath{\mathrm{{\,s}}}\xspace}

\def\ns   {\ensuremath{{\mathrm{ \,ns}}}\xspace}
\def\ps   {\ensuremath{{\mathrm{ \,ps}}}\xspace}

\def\mhz  {\ensuremath{{\mathrm{ \,MHz}}}\xspace}

\def\neutroneq {\ensuremath{\mathrm{ \,n_{eq}}}\xspace}

\def\gsim{{~\raise.15em\hbox{$>$}\kern-.85em
          \lower.35em\hbox{$\sim$}~}\xspace}
\def\lsim{{~\raise.15em\hbox{$<$}\kern-.85em
          \lower.35em\hbox{$\sim$}~}\xspace}

\def\degrees{\ensuremath{^{\circ}}\xspace}

\def\tell1  {TELL1\xspace}
\def\ukl1   {UKL1\xspace}

\usepackage{makecell}
\usepackage{multirow}
\def\tab {Table~}
\def\fig {Figure~}

\def\sect {Section~}

\def\app {Appendix~}
\def\eq {Eq.~}
\def\ttt {time-to-threshold\xspace}
\def\nonirr {nonirradiated\xspace}
\def\np {{n-on-p}\xspace}
\def\nn {{n-on-n}\xspace}

\def\maxfluence {\ensuremath{8 \times 10^{15} 1 \mev \neutroneq {\mathrm{ \,cm}}^{-2}}\xspace} 
\def\fluence {\ensuremath{{10^{15}  1 \mev \neutroneq {\mathrm{ \,cm}}^{-2}}}\xspace} 
 
\usepackage{booktabs}
\usepackage{setspace}
\usepackage{placeins}
\title{Temporal characterisation of silicon sensors on Timepix3 ASICs}

\author[a,d,1]{E.~Dall'Occo,\note{Corresponding author.}}
\author[a]{K.~Akiba,}
\author[a]{M.~van Beuzekom,}
\author[b]{E.~Buchanan,}
\author[c]{P.~Collins,}
\author[c]{T.~Evans,}
\author[e]{V.~Franco Lima,}
\author[a]{R.~Geertsema,}
\author[c]{H.~Schindler,}
\author[a]{H.~Snoek,}
\author[a,f]{and P.~Tsopelas}

\affiliation[a]{Nikhef, Science Park 105, 1098 XG Amsterdam, the Netherlands}
\affiliation[b]{University of Bristol, Beacon House,\\Queens Road, BS8 1QU, Bristol, United Kingdom}
\affiliation[c]{CERN, 1211 Geneve, Switzerland}
\affiliation[d]{Now at TU Dortmund,\\Otto-Hahn-Straße 4, 44227 Dortmund, Germany}
\affiliation[e]{Oliver Lodge Laboratory, University of Liverpool, \\Liverpool, L69 7ZE, United Kingdom}
\affiliation[f]{Now at Spectricon, Science and Technology Park of Crete, Heraklion, Greece}

\emailAdd{elena.dall'occo@cern.ch}

\abstract{
The timing performance of silicon sensors bump-bonded to Timepix3 ASICs is investigated, prior to and after different types of irradiation up to \maxfluence.
The sensors have been tested with a beam of charged particles in two different configurations, perpendicular to and almost parallel to the incident beam.
The second approach, known as the grazing angles method, is shown to be a powerful method to investigate not only the charge collection, but also the \ttt properties as a function of the depth at which the charges are liberated.
}

\keywords{Radiation-hard detectors; Hybrid detectors; Solid state detectors; Particle tracking detectors (Solid-state detectors); Radiation damage to detector materials (solid state); Timing detectors}

\arxivnumber{2102.06088} % only if you have one

\begin{document}
\maketitle

%% Uncomment during review phase. 
%% Comment before a final submission.
%\linenumbers

% You can include short sections directly in the main tex file.
% However, for larger papers it is desirable to split the text into
% several semiautonomous files, which can be revised independently.
% This is especially useful when developing a document in
% collaboration with several people, since then different parts can be
% edited independently.  This type of file organization is shown here.
% 
%%%%%%%%%%%%%%%%%%%%%%%%%%%%%%%%%%%%%%%%%%%%%%%%%%%%%%%%%%%%%%%%%%%%%%%%%%%%%%%%%%%%%%%%
\section{Introduction}
\label{sec:intro}

Understanding the timing capabilities of silicon sensors is fundamental for operation in a high rate environment such as the LHC. 
In view of the upgrade of the LHCb VErtex LOcator (VELO)~\cite{LHCb-TDR-013-m} a wide range of prototype sensors have been tested in a beam of charged particles in order to assess their suitability.
The upgraded VELO is a hybrid silicon pixel detector capable of $40\mhz$ readout, which surrounds the proton-proton collision region and is dedicated to the tracking and reconstruction of primary and secondary vertices. 
The main challenge for the operation of the sensors is the high and nonuniform radiation exposure, with a maximum fluence of \maxfluence, expected at the closest point to the proton-proton collision after $50 \invfb$ of integrated luminosity.
Hence, the prototype sensors have been tested before and after different irradiation types and fluences.

This paper presents an investigation of the temporal properties of different sensor prototypes.
The timing performance is determined using a particle beam with two complementary methods: at normal incidence and with a grazing angle~\cite{chiochia2004,1748-0221-7-09-P09011} approach.
The grazing angle method consists of analysing particle tracks that traverse the sensor almost parallel to its surface, such that the depth that a charge is deposited at can be inferred from the position of the corresponding hit within a cluster.
The depletion depth of planar sensors can be precisely determined using the grazing angle technique ~\cite{Henrich:687041}.
It has also been applied in charge diffusion studies in silicon~\cite{7272141} and to perform intrinsic spatial resolution studies~\cite{1748-0221-8-11-P11007}.
In this paper, the grazing angle technique has been further developed to study the time properties of the sensors as a function of depth.
A complementary technique for studying sensor properties as a function of depth is the transient current technique (TCT)~\cite{Eremin:1996yd,Kramberger:2010tem,Garcia:2017xap}, which provides a characterisation and visualisation of the electric field distribution.

Two main figures-of-merit are used to quantify the timing performance: the time from when the charge is liberated to when the signal crosses the threshold on average, referred to as \ttt, and the width of the \ttt distribution, referred to as the temporal resolution.

This paper is organised as follows.
In \sect\ref{sec:assemblies} the sensor prototypes are described, followed by the experimental setup in \sect\ref{sec:telescope}.
In \sect\ref{sec:method} the \ttt measurement method is discussed.
The results for nonirradiated, uniformly neutron irradiated and nonuniformly proton irradiated sensors are presented in \sect\ref{sec:perptiming} and \sect\ref{sec:grazing} for sensors placed perpendicularly to the beam and at grazing angles, respectively.
The conclusions are drawn in \sect\ref{sec:conclusion}.

%%%%%%%%%%%%%%%%%%%%%%%%%%%%%%%%%%%%%%%%%%%%%%%%%%%%%%%%%%%%%%%%%%%%%%%%%%%%%%%%%%%%%%%%
\section{Methodology}
\label{sec:methodology}
\subsection{Detector assemblies under study}
\label{sec:assemblies}

The prototype assemblies tested are hybrid pixel detectors, composed of sensors with $256\times256$ pixels of 55\mum pitch bump-bonded to Timepix3 ASICs~\cite{poikela:timepix3}.
The assemblies are glued to a $635 \mum$ thick AlN ceramic board and wire-bonded to a custom made kapton-copper hybrid, also glued to the ceramic substrate,
subsequently read out by a SPIDR system~\cite{Visser:2015bsa}.

The prototype sensors were produced by two different manufacturers, Hamamatsu~Photonics~K.~K. (HPK)\footnote{Hamamatsu Photonics K. K., 325-6, Sunayama-cho, Naka-ku, Hamamatsu City, Shizuoka, 430-8587, Japan} and Micron Semiconductor Ltd\footnote{Micron Semiconductor Ltd, 1 Royal Buildings, Marlborough Road, Lancing BN158UN, United Kingdom}.
The prototype sensors have different design features, such as pixel implant size, sensor thickness, bulk type, pixel-to-edge (PTE) distance. 
The main characteristics of the assemblies are summarised in \tab\ref{tab:assemblies}.
The details of individual assemblies and their identification numbers, which will be used in the following, can be found in \app\ref{app:assemblies}.
\begin{table}[!h]
\caption{Prototype assemblies.}
\label{tab:assemblies}
\centering
\begin{tabular}{llccc}
  \toprule
Vendor & Type & Thickness & PTE & Implant width \\
\midrule
HPK    & \np & 200\mum & 450, 600\mum & 35, 39\mum \\
Micron & \np & 200\mum & 450\mum & 36\mum   \\
Micron & \nn & 150\mum & 250, 450\mum & 36\mum   \\
\bottomrule
\end{tabular}
\end{table}

The Timepix3 ASIC can simultaneously measure the threshold crossing time, denoted by Time-of-Arrival (ToA), and the time the signal is above threshold, denoted by Time-over-Threshold (ToT).
The former is registered with a Time-to-Digital Converter (TDC) with a bin width of 1.56~\ns. 
The latter is related to the energy deposited and is converted into equivalent units of collected electrons via a charge calibration process, as described in Ref.~\cite{Akiba:2019faz}.
More details on the Timepix3 ASIC can be found in \tab\ref{tab:tpx3}.
The threshold is set at 1000~\en to ensure the data is free from noise. 
\begin{table}[t]
\caption{Characteristics of the Timepix3 ASIC.}
\label{tab:tpx3}
\centering
\begin{tabular}{ll}
  \toprule
  technology & $130 \nm$ CMOS \\
  peaking time & $ < 25 \ns$ \\
  noise & 80 - 100~\en \\
  TDC bin width & 1.56~\ns \\
\bottomrule
\end{tabular}
\end{table}
\begin{table}[b!]
\caption{Characteristics of the  facilities used for the irradiation programme.}
\label{tab:irrfacilities}
\centering
\begin{tabular}{lccccc}
  \toprule
  Facility   & Particles             & Cooling & Scanning  &   \makecell{Intensity \\  {[$10^{12}$\,s$^{-1}$\,cm$^{-2}$]} } & \makecell{Hardness \\ {Factor}} \\
  \midrule
  IRRAD      & 24\,\gev\ protons   & yes     & no        & 0.02                                        & $0.62\pm0.04$ \cite{Allport_2019} \\
  Ljubljana  & reactor  neutrons  & no      & no        & 3.05                                        &$0.90\pm0.03$ \cite{Kramberger:1390490} \\
\bottomrule
\end{tabular}
\end{table}
\begin{figure}[b!]
    \centering
    {\includegraphics[width=0.49\textwidth]{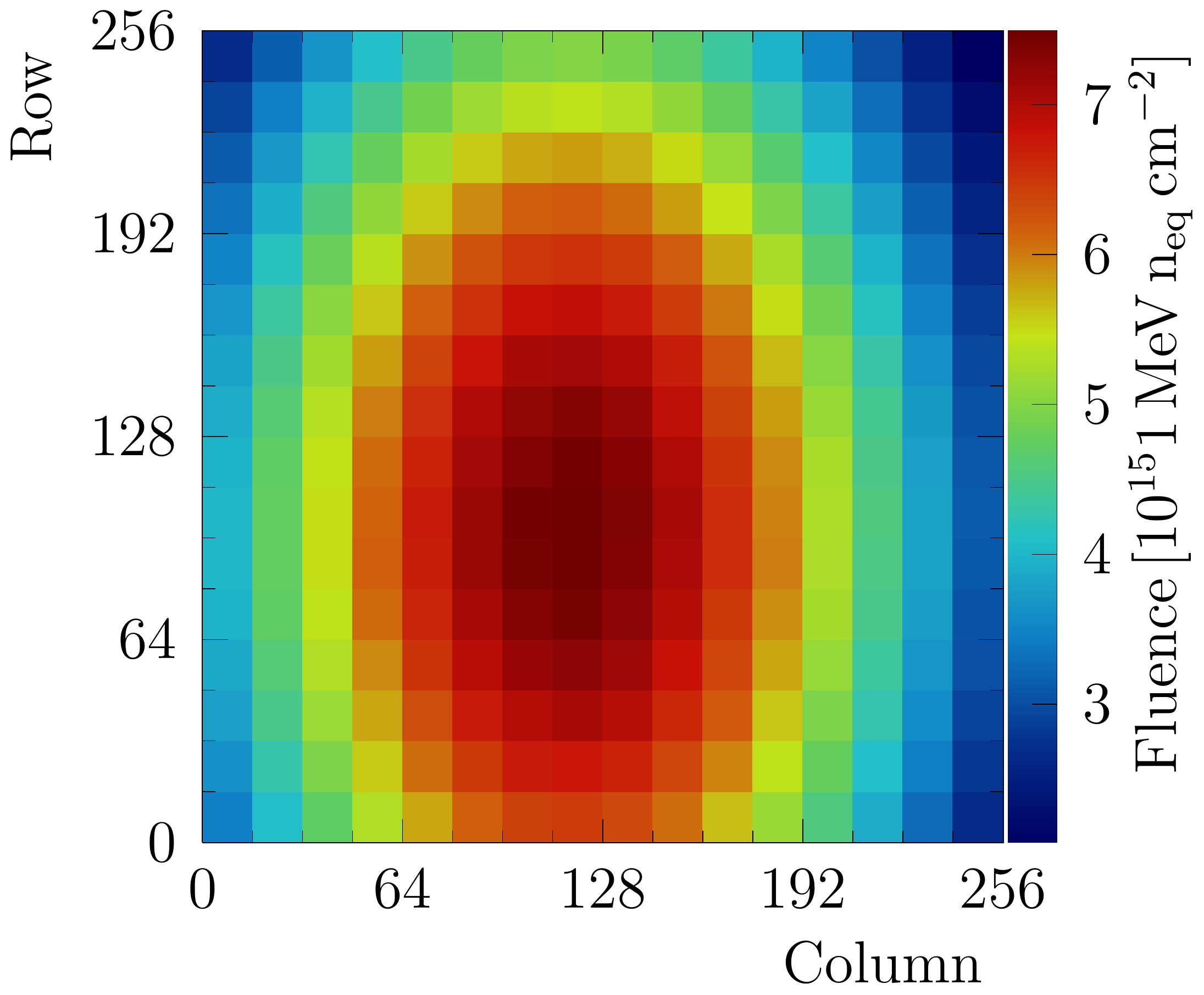}}
    \caption{Reconstructed fluence profile.}
  \label{fig:fluence}
\end{figure}

Several assemblies were irradiated up to the maximum fluence of \maxfluence, with some exposed to $24\gevc$ protons at IRRAD and others to neutrons at the JSI reactor in Ljubljana.
The uncertainty on the fluence at both facilities is estimated to be of the order of 10\%. 
The main characteristics of the two facilities have been summarised in \tab\ref{tab:irrfacilities}.
During the shipping from the irradiation facilities the sensors were kept at room temperature for 11 days, while in between measurements they were stored in a freezer at a temperature of $-15^{\circ}\text{ C}$ to reduce possible annealing effects.
Both types of irradiation have been investigated as the damage to the sensor is different~\cite{Moll:2640820} and the actual radiation environment of the detector will be a mixture of the two.
A nonuniform irradiation profile is used at IRRAD to emulate the expected conditions of the upgraded VELO.
The irradiation profile is reconstructed combining the activation map of the assembly and the measurement provided by the dosimetry survey.
This method is thoroughly described in Ref.~\cite{DallOcco:2020nir}. 
The reconstructed fluence profile is shown in \fig\ref{fig:fluence}.
%

%%%%%%%%%%%%%%%%%%%%%%%%%%%%%%%%%%%%%%%%%%%%%%%%%%%%%%%%%%%%%%%%%%%%%%%%%%%%%%%%%%%%%%%%
\subsection{Test beam campaign}
\label{sec:telescope}

An extensive test beam programme has been carried out at the SPS H8 beamline at CERN to characterise the sensors.  
The beam is a mixed charged hadron beam 
($\sim 67\%$ protons, $\sim30\%$ pions) at $180\gevc$. 
The trajectories of particles are reconstructed with the Timepix3 telescope~\cite{Akiba:2019faz}, a high rate, data-driven beam telescope, composed of two arms of four planes each.  
Each plane is instrumented with a 300\mum p-on-n silicon sensor bump-bonded to a Timepix3 ASIC. 
The centre of the telescope is reserved for the Device Under Test (DUT). 
The DUT area is equipped with remotely controlled motion stages able 
to translate the DUT in $x$ and $y$ directions (orthogonal to the beam axis) 
and to rotate it about the $y$ axis. 
A vacuum box can also be installed on the central stage  to facilitate 
testing of irradiated devices at high voltage. The DUT can be cooled down 
to temperatures of about $ -20^{\circ}$C.

The pointing resolution at the DUT position is about $1.6\mum$, 
enabling intrapixel studies of the sensor.
The typical temporal resolution on a track using only timestamps of the telescope Timepix3 planes is about $350 \ps$. 
In the grazing angle configuration, an excellent temporal resolution is useful as clusters can be associated to tracks only using the timing information, which avoids the complexities of performing the association with only one spatial dimension. 
 
%%%%%%%%%%%%%%%%%%%%%%%%%%%%%%%%%%%%%%%%%%%%%%%%%%%%%%%%%%%%%%%%%%%%%%%%%%%%%%%%%%%%%%%%
\subsection{Time-to-threshold measurement}
\label{sec:method}

\begin{figure}[t!]
    \centering
  {\includegraphics[width=0.49\textwidth]{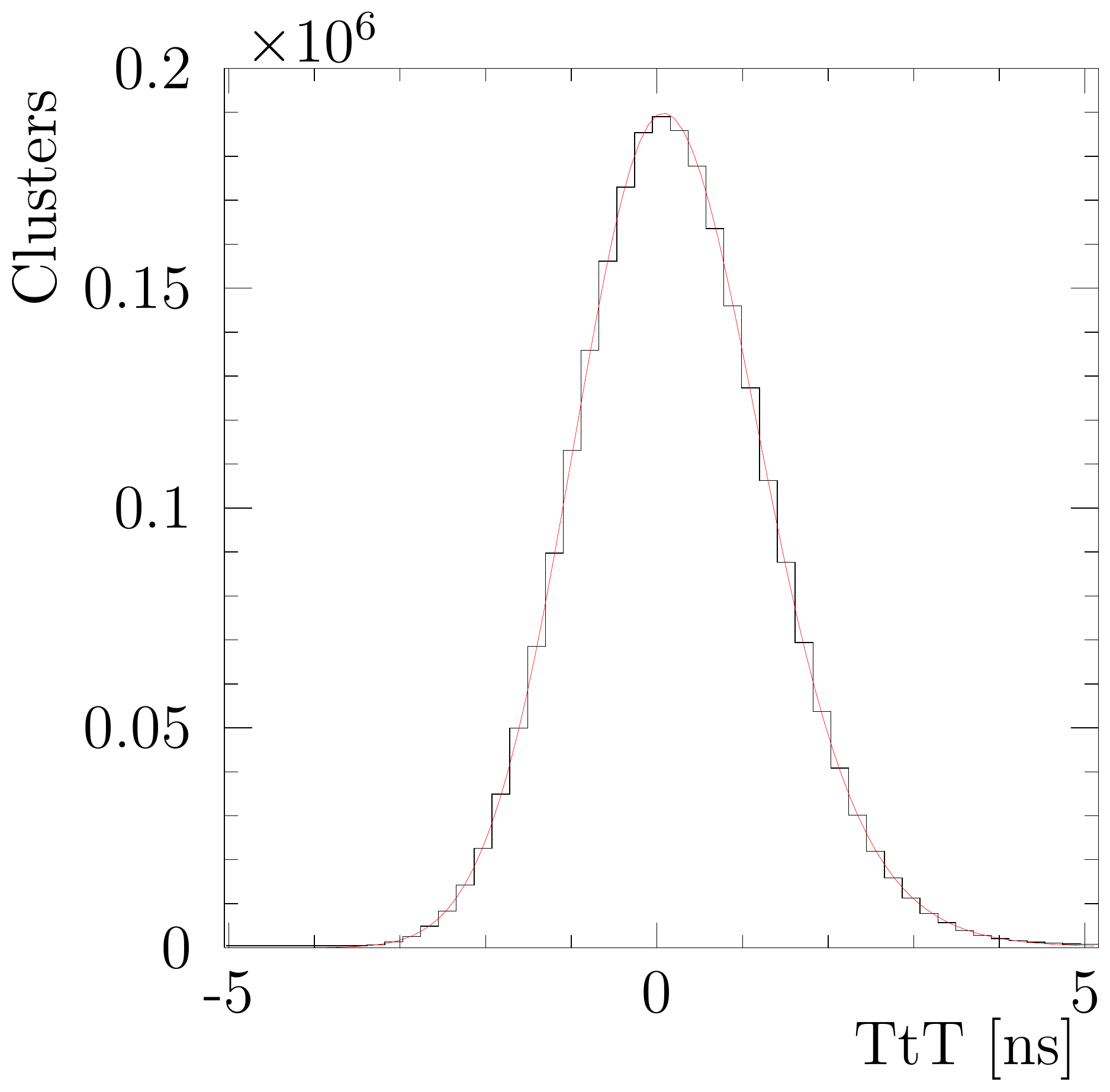}}\hspace{1mm}
  {\includegraphics[width=0.49\textwidth]{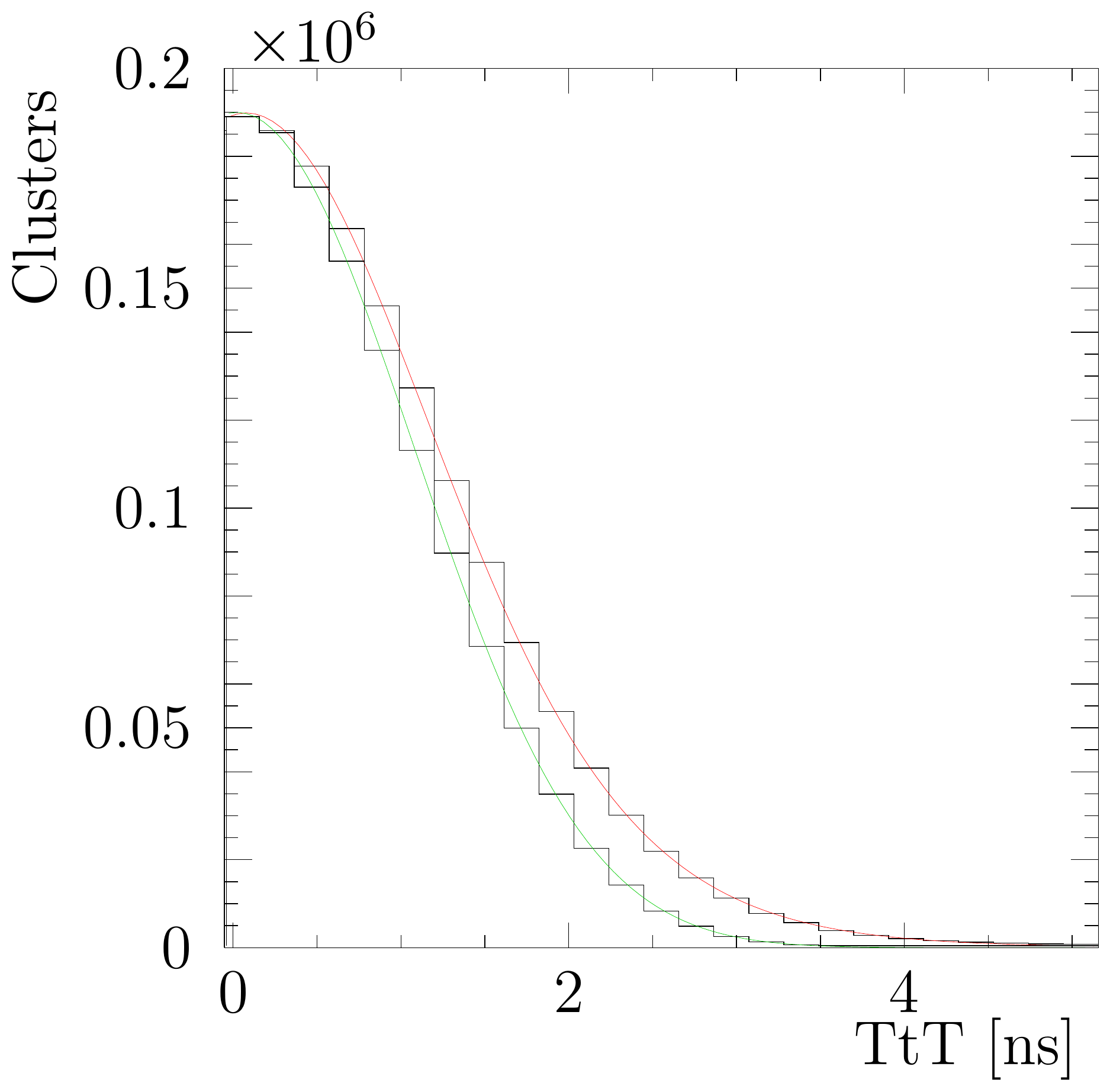}}  
    \caption{Example of a typical \ttt (TtT) distribution for a sensor irradiated to full fluence after timewalk correction (left), and the same distribution mirrored around the y-axis (right), where the right-hand side is indicated in red and the left-hand side in green.}
  \label{fig:TtTexample}
\end{figure}

The \ttt~(TtT) of a hit is obtained by subtracting the track time provided by the telescope from the hit time measured in the DUT.
There is a constant offset between the track and DUT hit times due to time-of-flight and differences in cable length between the DUT and the telescope planes, and hence only relative differences in the \ttt are meaningful.
The most probable value of the \ttt is determined by fitting the distribution with a Cruijff~\cite{delAmoSanchez:2010ae} function, a Gaussian function with different left-right widths and non-Gaussian tails
\begin{equation}
  f(x;x_0,\sigma_L,\sigma_R,\alpha_L,\alpha_R) = \begin{cases}
    \exp \big( - \frac{(x-x_0)^2}{2(\sigma_L^2+\alpha_L(x-x_0)^2)}\big), & \text{if } x < x_0,\\ 
    \exp \big( - \frac{(x-x_0)^2}{2(\sigma_R^2+\alpha_R(x-x_0)^2)}\big), & \text{if } x > x_0,
  \end{cases}
\end{equation}
where $x_0$ is the mean, $\sigma_{L,R}$ is the left-right width and $\alpha_{L,R}$ parametrises the left-right tail.
The right width is in general larger than the left width due to timewalk.
An example of the fit to the \ttt distribution is shown in \fig\ref{fig:TtTexample}.
Unless otherwise stated, the quoted resolution corresponds to the right width of the Cruijff function.
The measured resolution is the sum in quadrature of the intrinsic resolution of the DUT and the resolution of the telescope, of $ 350 \ps$~\cite{Akiba:2019faz}.
The resolution of the DUT can be described by a combination of three terms~\cite{Cartiglia_2014}
\begin{equation}
\label{eq:treso}
    \sigma_t^2 = \left( \left[ \frac{t_r V_{\text{th}}}{S} \right]_{\text{RMS}} \right)^2 + \left( \frac{t_r}{S/N} \right)^2 + \left( \frac{\text{TDC}_{\text{bin}}}{\sqrt{12}} \right)^2,
\end{equation}
where the first component is the contribution from timewalk, the second component is the contribution from jitter, and the last component is the contribution from TDC binning. 
Here $t_r$ is the rise time of the signal at the output of the amplifier, $V_{\text{th}}$ is the threshold of the discriminator, $S$ is the amplitude of the signal, $N$ is the noise of the front-end, and $\text{TDC}_{\text{bin}}$ is the TDC bin width. 
The contribution due to Landau fluctuations is estimated to be lower than $50 \ps$ for both nonirradiated and irradiated sensors, and hence negligible compared to the other terms.
The additional contribution from different time offsets within the pixel matrix is consistent between all ASICs~\cite{Heijhoff_2020} and thus is neglected.

In order to understand the charge collection time, the effects of the sensor must be disentangled from those of the ASIC. 
It is particularly important to correct for timewalk after irradiation due to the degraded charge collection.
For studies at normal incidence, the timewalk curve is determined per pixel by injecting a test pulse with known charge in the pixel front-end. 
Conversely, the grazing angle method has the advantage that the timewalk curve can be determined directly from the testbeam data, by selecting only charges liberated at small depth, up to about $25\mum$ from the pixel electrodes, in order to reduce the effect of the sensor to a negligible level.
The timewalk curve obtained with this method is found to be compatible with that obtained from test pulse data for a subset of the assemblies, an example of which is shown in \app\ref{app:timewalk}.

The timewalk curve is parameterised as
\begin{equation}
  t(q) = \frac{A}{q-q_0} + C,
\end{equation}
where $t$ is the \ttt, $q$ is the charge, $q_0$ the charge corresponding to the onset of the asymptote, $A$ the slope and $C$ the offset.
The inverse function is used to correct the measured \ttt of each hit.
Unless otherwise specified, the results shown are corrected for timewalk. 

%%%%%%%%%%%%%%%%%%%%%%%%%%%%%%%%%%%%%%%%%%%%%%%%%%%%%%%%%%%%%%%%%%%%%%%%%%%%%%%%%%%%%%%%
\section{Results at perpendicular incidence}
\label{sec:perptiming}

In this section, the time response of the different sensor designs is studied prior to and after irradiation, with some prototypes exposed to uniform and others to nonuniform irradiation profiles. 
The prototypes are placed perpendicular to the incident beam, thus the charge is liberated along the thickness of the sensor allowing for a direct measurement of the resolution per pixel.

\subsection{Nonirradiated sensors}
\begin{figure}[b]
  \centering
  {\includegraphics[width=0.49\textwidth]{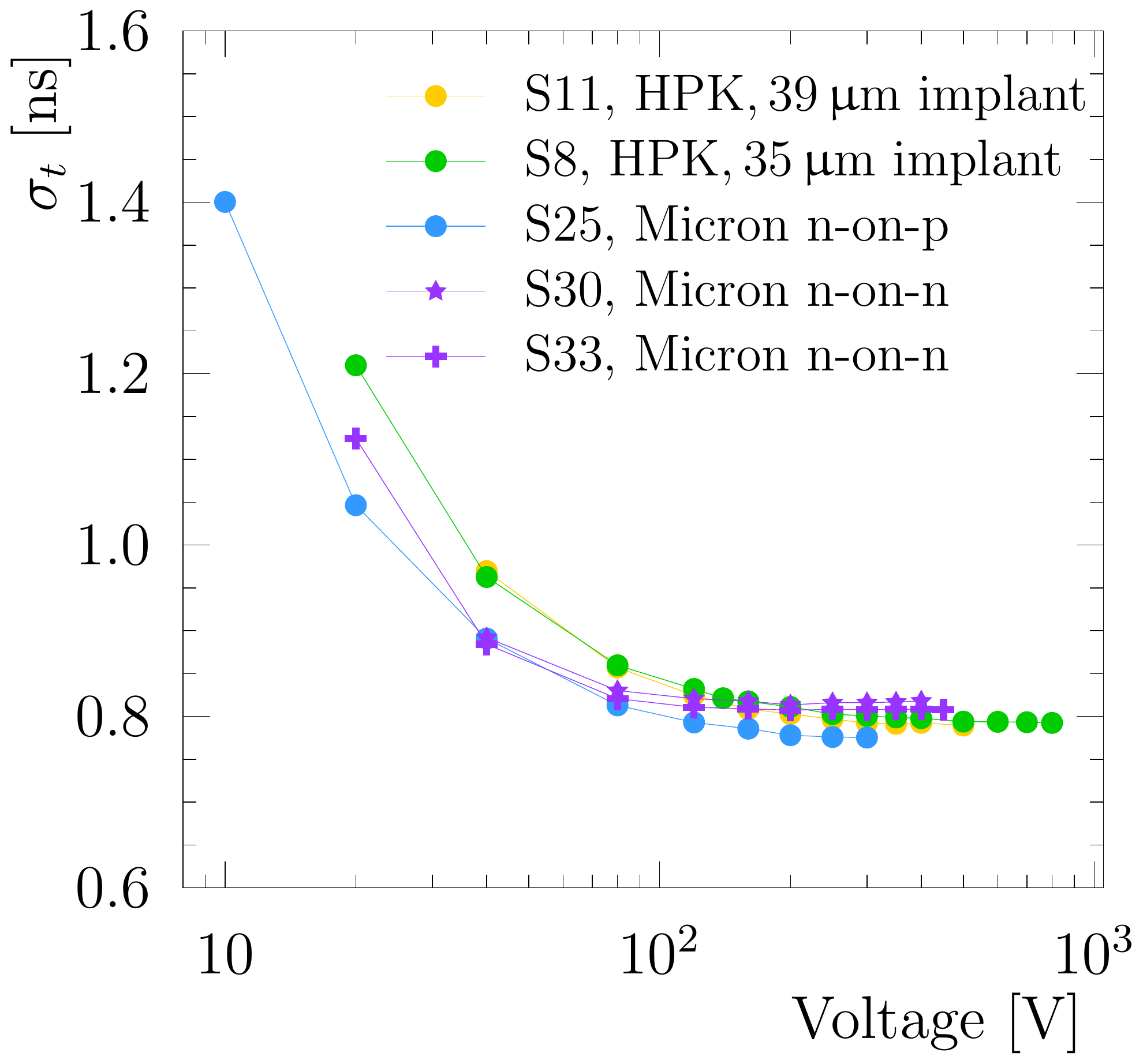}} \hspace{1mm}
  {\includegraphics[width=0.49\textwidth]{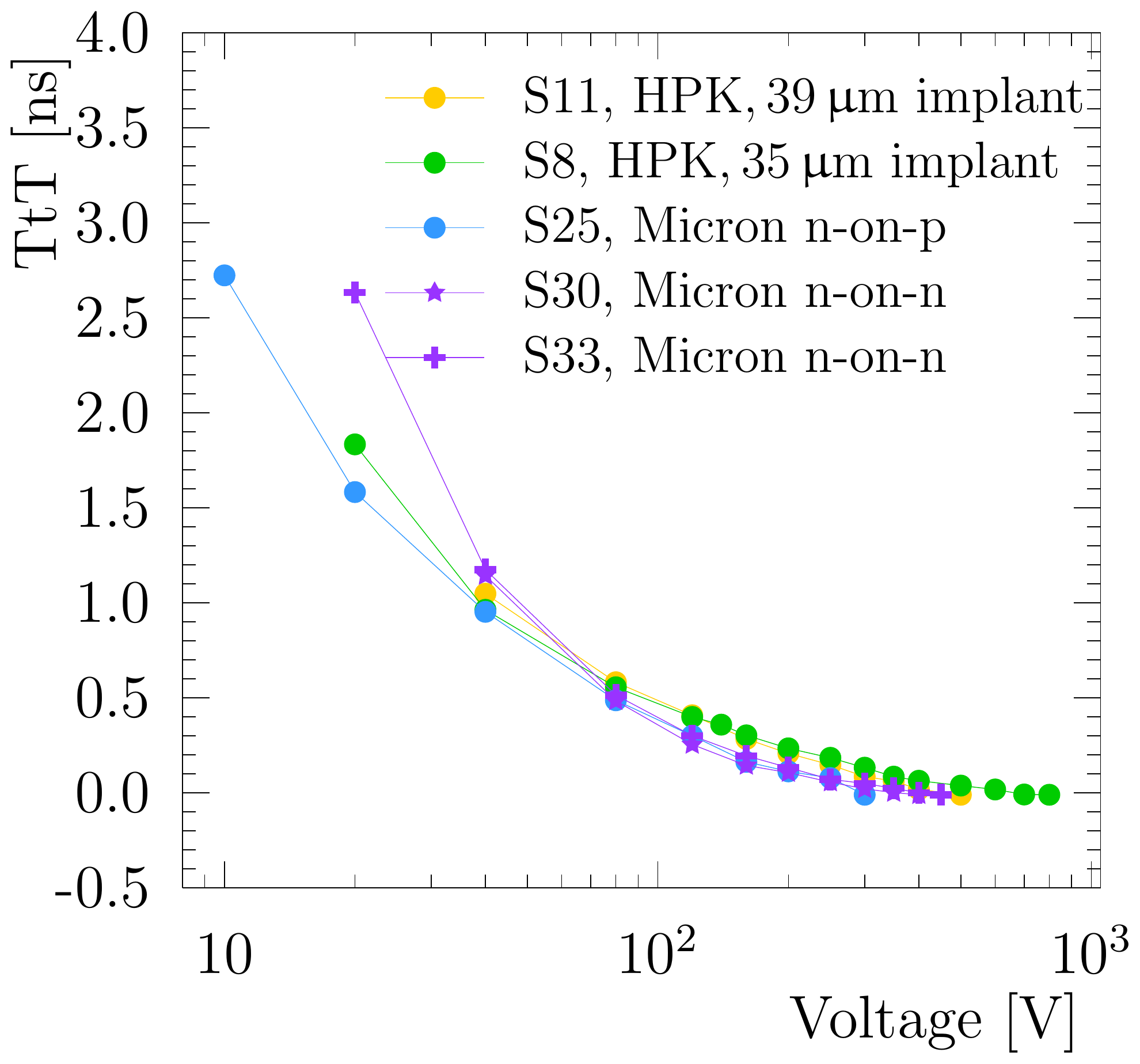}} 
  \caption{Resolution (left) and \ttt (right) as a function of operating voltage for different nonirradiated sensors. 
  }
  \label{fig:timeResBefore}
\end{figure}
\begin{figure}[!b]
    \centering
  {\includegraphics[width=0.49\textwidth]{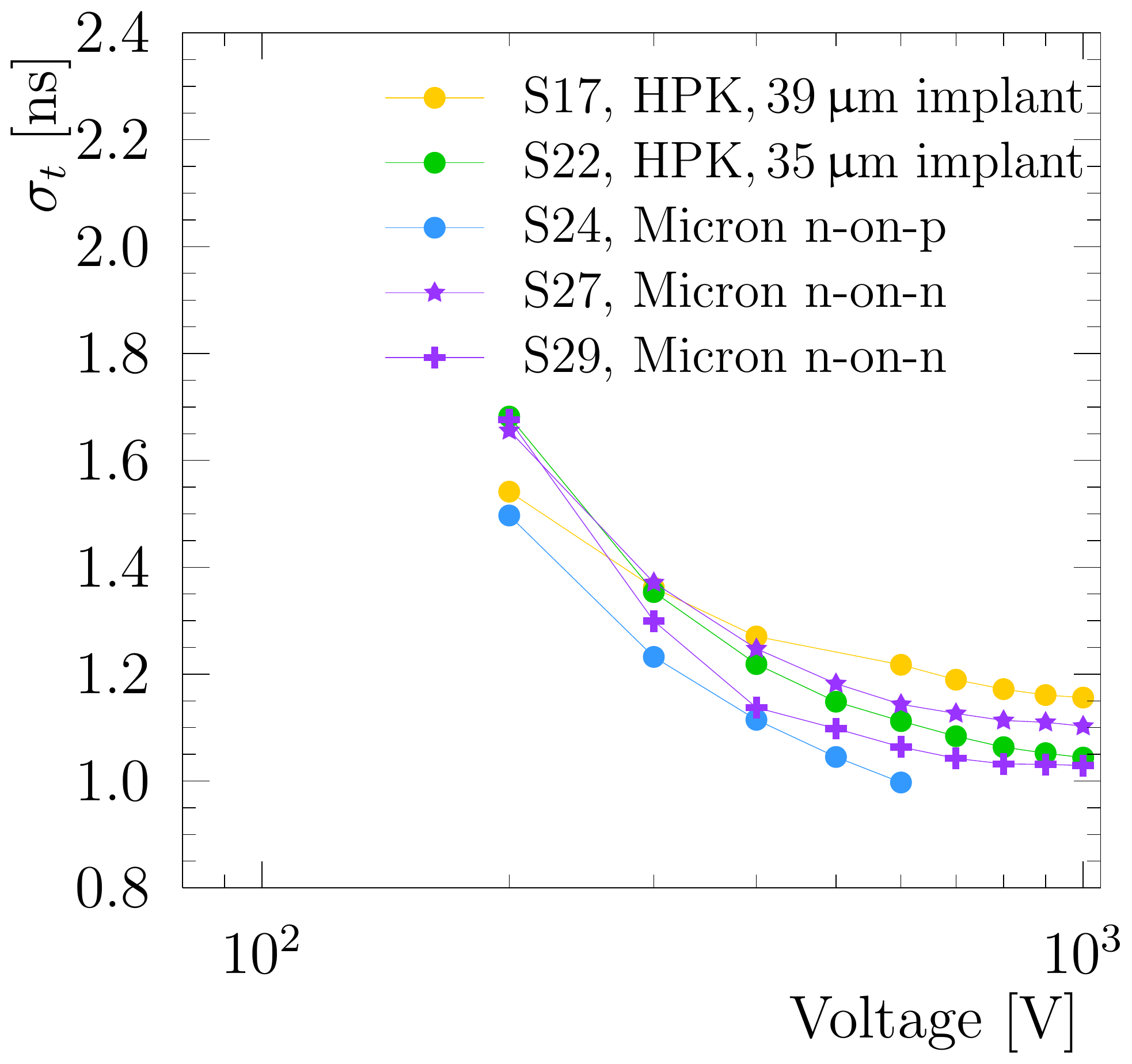}}\hspace{1mm}
  {\includegraphics[width=0.49\textwidth]{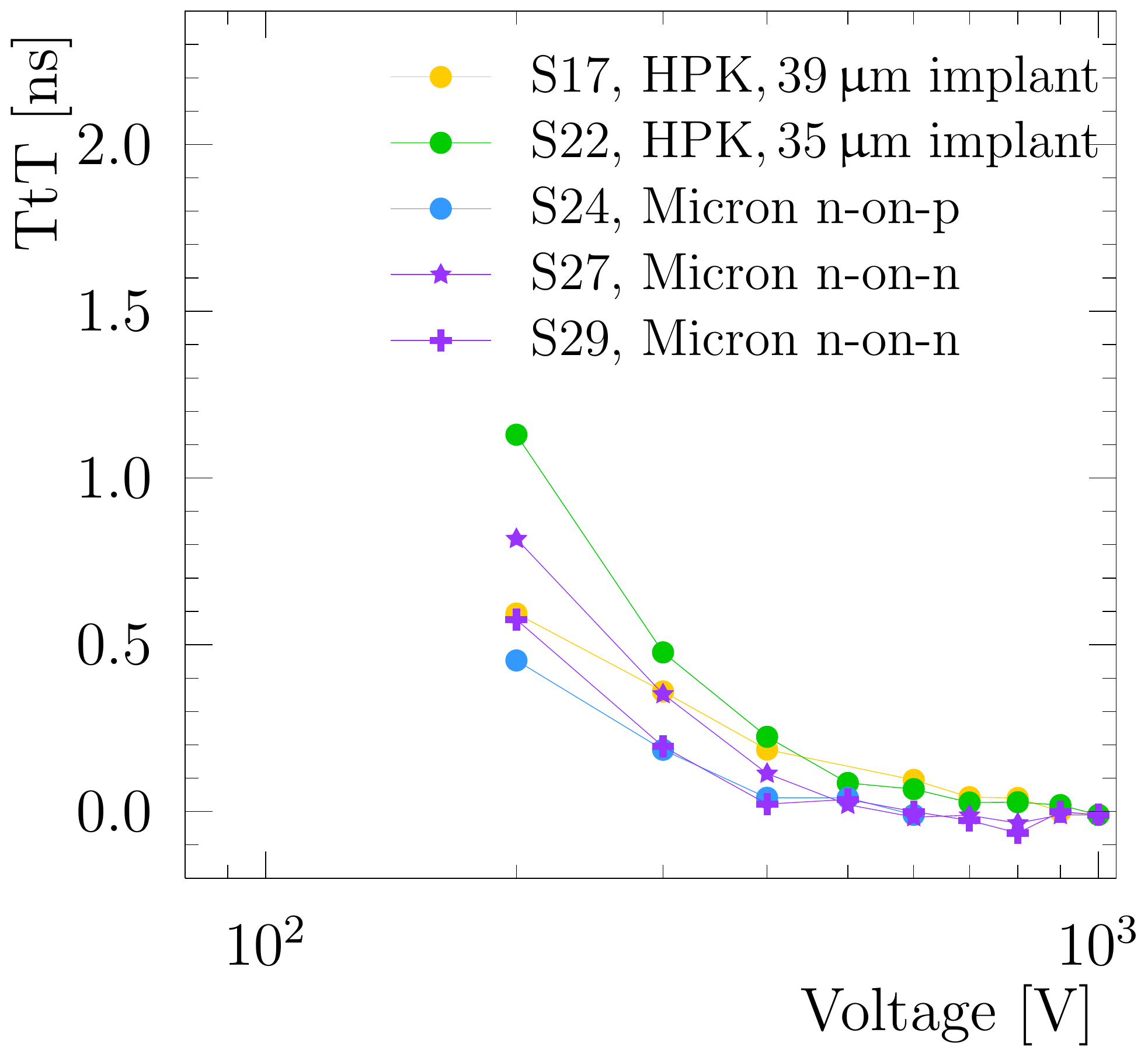}} 
  \caption{Resolution (left) and \ttt (right) as a function of operating voltage for sensors irradiated to full fluence at JSI. 
  }
  \label{fig:TimeResUni}
\end{figure}
Five assemblies have been tested prior to irradiation in order to disentangle sensor effects from those caused by radiation damage.
For these sensors, the resolution and the \ttt are shown as a function of the operating voltage in \fig\ref{fig:timeResBefore}.
The resolution for all the sensor types improves with increasing voltage and saturates at around 0.8~ns. 
At the highest voltages, the resolution is dominated by contributions from the ASIC, with comparable contributions from jitter, TDC binning and different time offsets within the pixel matrix~\cite{Heijhoff_2020}.
Therefore the resolution saturates at lower voltage than the \ttt.
The \ttt of the signal decreases with the operation voltage indicating that the electric field is increasing and the charge carriers do not reach their saturation velocity until a voltage of around 1000~V. 
All types of sensors achieve a similar temporal resolution and \ttt at high voltage, while a small difference arises between the different types at lower voltages.

\subsection{Uniformly irradiated sensors}

The resolution for irradiated sensors is expected to be worse due to an increased jitter in the signal. 
After irradiation the amount of charge collected decreases, degrading the signal-to-noise ratio and thus the resolution.
The mobility of the charge carriers also decreases due to radiation damage~\cite{Borchi:1994qw,Leroy_2007}, further degrading the time resolution. 

The resolution for sensors irradiated to \maxfluence are shown in \fig\ref{fig:TimeResUni} (left).
All the sensors show the same trend, with the resolution improving as the applied bias increases.
This can mainly be attributed to the change in signal amplitude, which increases linearly with applied bias from 200 to 1000~V~\cite{Geertsema:2020pmc} with no evidence of saturation.
The resolution is also not seen to saturate at 1000~V, indicating that better timing performance could potentially be achieved by further increasing the bias voltage.
\fig\ref{fig:TimeResUni} (right) shows that the \ttt trend as a function of voltage is similar between the different types of sensor after irradiation.
The \ttt is worse than prior to irradiation, especially at a bias voltage below 400~V. 

It can be concluded that after irradiation to the maximum fluence, the sensors would need to be operated with a voltage of at least 1000~V to optimise the resolution. 
Such a high voltage also mitigates the effect of timewalk, which is beneficial for the operations of the detector by reducing the probability that hits are recorded in the subsequent bunch crossing. 

\begin{figure}[t]
  \centering
  {\includegraphics[width=0.49\textwidth]{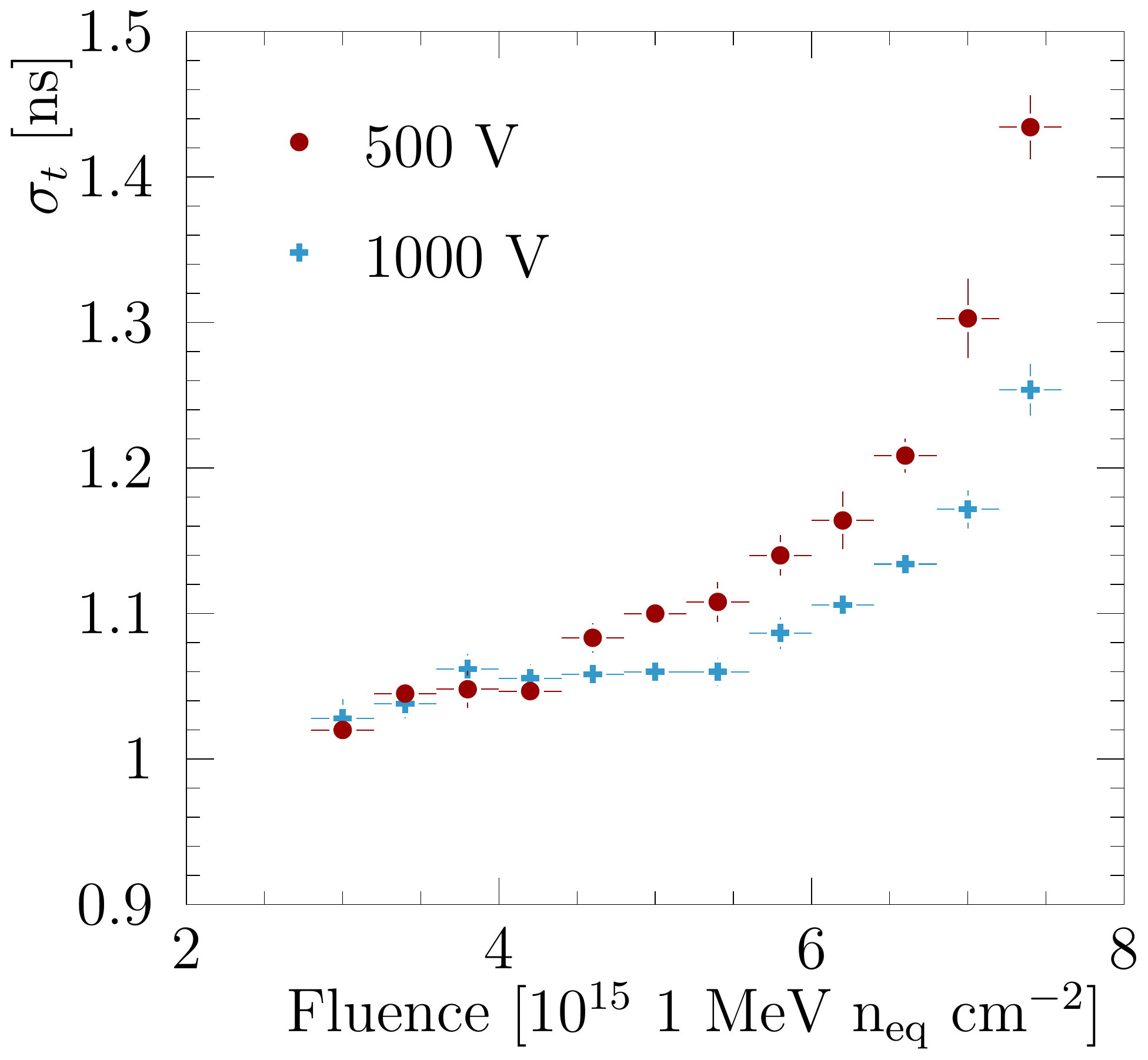}} \hspace{1mm}
 {\includegraphics[width=0.49\textwidth]{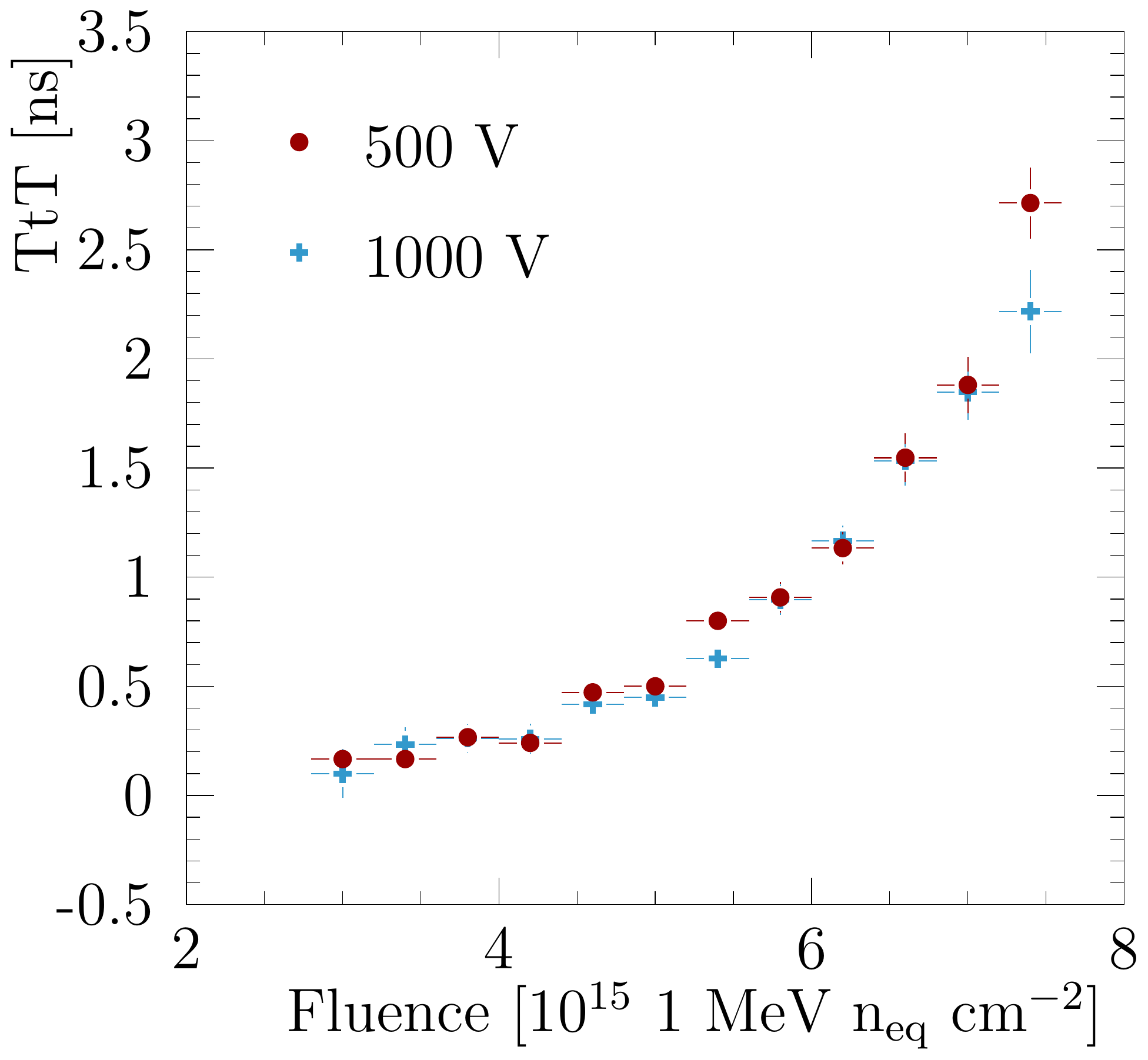}} \\
 \caption{Resolution (left) and \ttt (right) as a function of fluence for different operating voltages for a $200 \mum$ nonuniformly irradiated HPK \np sensor (S8).
  }
  \label{fig:TimeResS8}
\end{figure}
\begin{figure}[t]
  \centering
  {\includegraphics[width=0.49\textwidth]{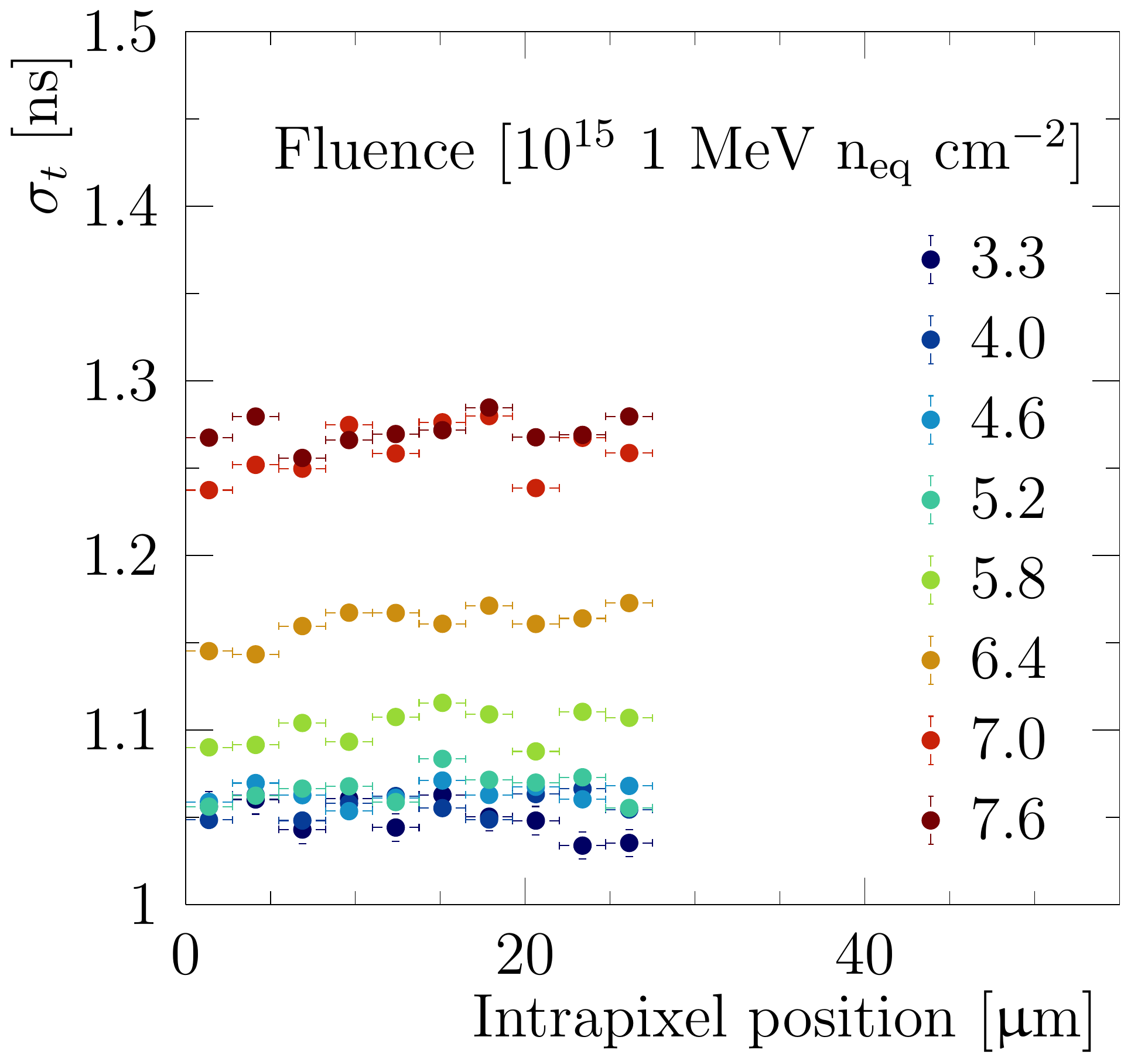}} \hspace{1mm}
 {\includegraphics[width=0.49\textwidth]{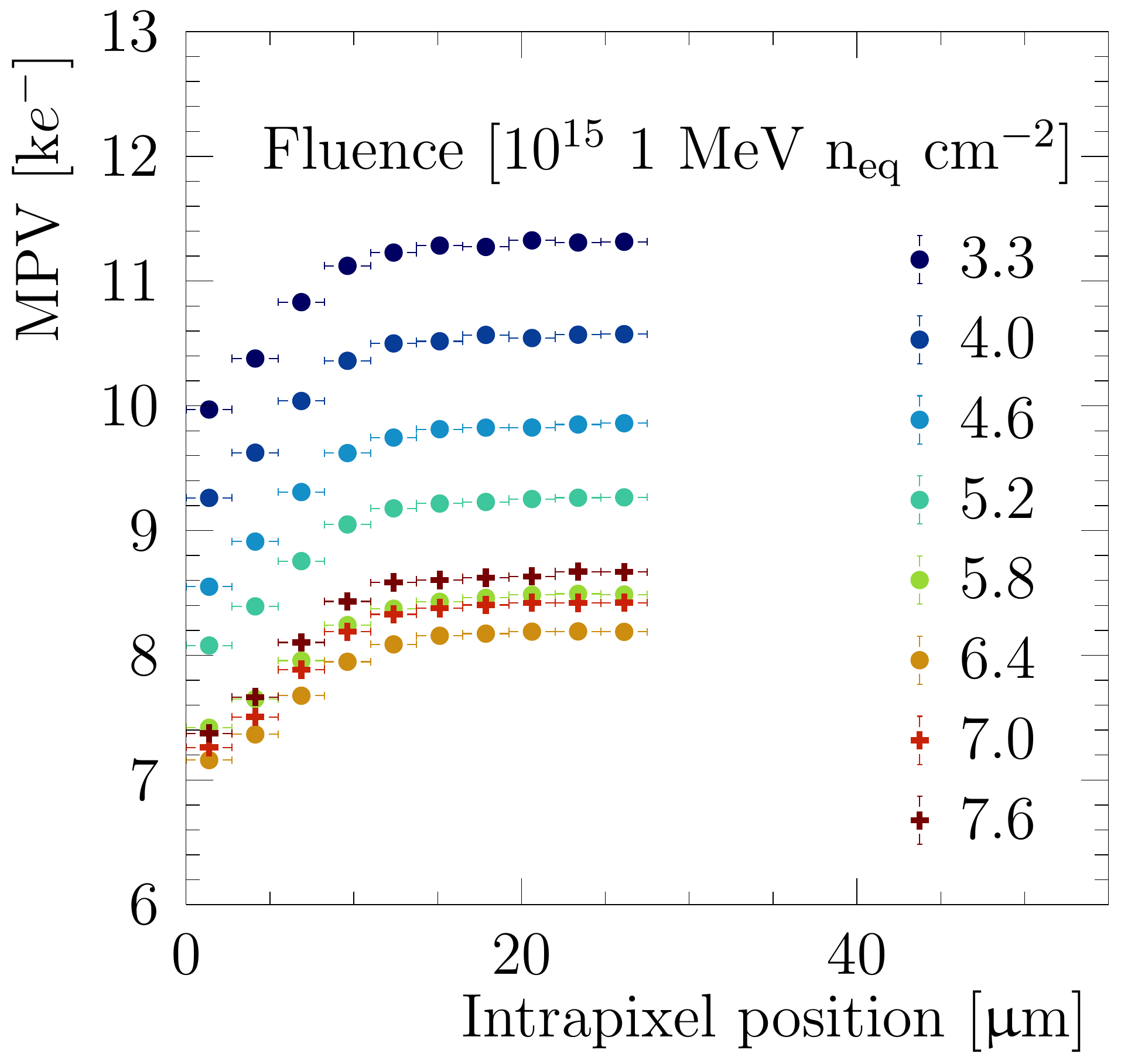}} \\
 \caption{Resolution (left) and MPV of the cluster charge (right) for a slice of $5\mum$ centred at $y=27.5 \mum$ of the pixel as a function of intrapixel position for different fluences for a $200 \mum$ nonuniformly irradiated HPK \np sensor (S8) operated at 1000~V. The uncertainties are statistical only.
  }
  \label{fig:intraTimeRes1}
\end{figure}

\subsection{Nonuniformly irradiated sensors}

The resolution is studied as a function of fluence using a nonuniformly irradiated sensor. 
\fig\ref{fig:TimeResS8} shows the resolution and \ttt of a HPK sensor for two different operating voltages, 500~V and 1000~V, with similar trends observed for intermediate voltages.
The resolution degrades with increasing fluence. 
While increasing the voltage improves the resolution in all cases, as was already noted in \fig\ref{fig:TimeResUni} (left), 
the performance achieved prior to irradiation is never attained. The \ttt for the two different operating voltages is comparable, as was already observed in \fig\ref{fig:TimeResUni} (right).

The variation of the resolution within a pixel for different fluences is investigated in order to understand if the observed degradation is localised.
The square pixel symmetry is exploited by combining the data from four quadrants into one in order to maximise the effective sample size for the intrapixel study.
The resolution and the \ttt of a slice of $5\mum$ centred at $y=27.5$\mum of the pixel, where the origin is the lower left corner, is shown in \fig\ref{fig:intraTimeRes1}. 
The resolution is uniform over the pixel for all fluence levels at an operating voltage of 1000~V.

Charge multiplication is observed for this sensor, as reported in Ref.~\cite{Geertsema:2020pmc}. 
The Most Probable Value (MPV) of the collected charge is presented in \fig\ref{fig:intraTimeRes1} in order to evaluate interplay of the timewalk and high fluence effects.
At the highest fluences, 7.0 and $7.6\times\fluence$,
the signal is larger than the immediate lower fluence bin, $6.4\times\fluence$.
Despite a higher signal due to charge multiplication, no improvement is observed in the resolution, being dominated by the ASIC contributions, nor \ttt.
%
%

%%%%%%%%%%%%%%%%%%%%%%%%%%%%%%%%%%%%%%%%%%%%%%%%%%%%%%%%%%%%%%%%%%%%%%%%%%%%%%%%%%%%%%%%
\section{Results at grazing angles}
\label{sec:grazing}

The grazing angle technique is used to study the \ttt as a function of the depth at which the charge is deposited in the sensor.
The charge collection as a function of depth is also measured, providing complementary information to understand the time dependent properties of the sensors. 
Firstly, the data selection and grazing angle method are described in \sect\ref{subsec:selection}, followed by the results for nonirradiated sensors in \sect\ref{subsec:grazingnonirr}, and uniformly neutron irradiated sensors and nonuniformly proton irradiated sensors in Sections~\ref{subsec:grazingunifirr} and \ref{subsec:grazingnonunifirr}, respectively.

\subsection{Data Selection}
\label{subsec:selection}

\begin{figure}[b!]
  \centering
  {\includegraphics[width=0.6\textwidth]{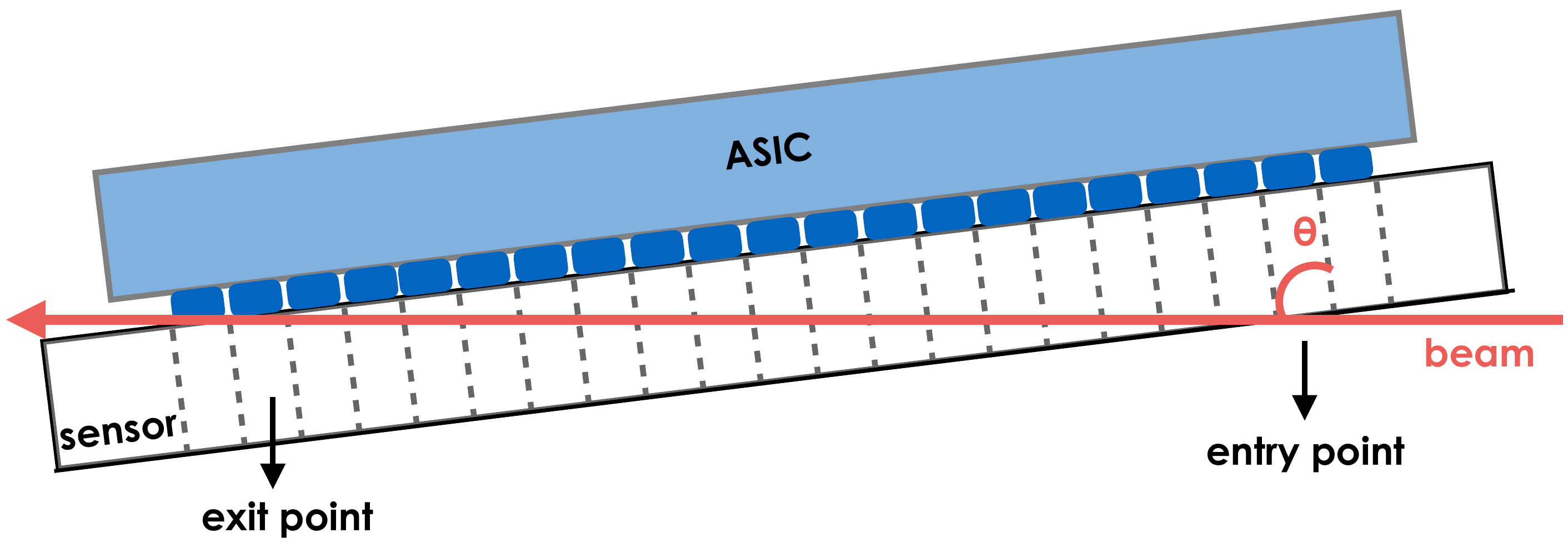}}\hspace{1mm}
  {\includegraphics[width=0.35\textwidth]{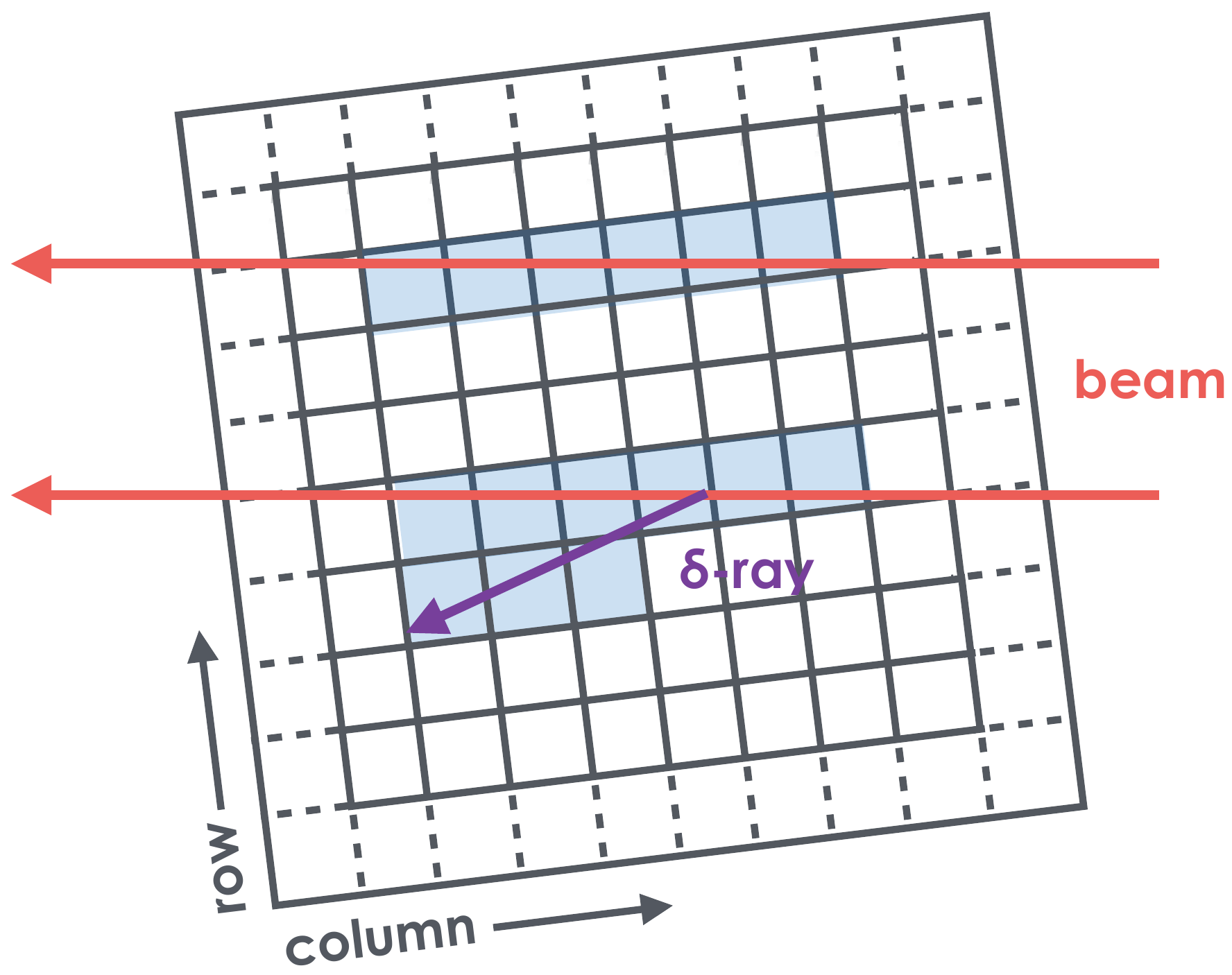}}
\caption{Illustration of the grazing angle setup.
Top view of the sensor (left):  the entry and exit point of the track are indicated, as well as the angle $\theta$ of the track with respect to the sensor.
Front view of the sensor (right): two different types of track are represented, with and without the emission of a $\delta$-ray.}
  \label{fig:scheme}
\end{figure}

In the grazing angle setup, the sensor is placed almost parallel to the beam such that the incident particle traverses multiple adjacent pixels, as illustrated in  \fig\ref{fig:scheme}.
Only clusters on the DUT that are associated to tracks reconstructed by the telescope are used.
A cluster is considered associated to a track if the time difference between the two is less than $10 \ns$, where the time of the cluster is defined as the earliest timestamp amongst its constituent pixel hits.
Given the average rate at the CERN SPS of 2M tracks for a spill of $4.5\sec$, the time cut is sufficient to perform the cluster-track association, expecting two tracks in the same time window in less than 0.5\% of the cases. 
Clusters are removed from the analysis in the case where the corresponding track could be associated to multiple clusters.
The cluster is required to span only one pixel row in order to remove particle tracks with delta-ray emissions or secondary particle production, as illustrated in \fig\ref{fig:scheme} (right), discarding about half of the data.
A gap of up to three contiguous empty pixels is allowed within the cluster to account for signals in dead or masked pixels, or with charge lower than the threshold in the case of irradiated sensors.
The entry and exit point of the track is required to be further than three pixels from the edge of the pixel matrix to ensure that the full cluster is within the sensitive volume.
The cluster length, which is given by the number of adjacent columns, depends on the incident angle $\theta$.
The expected cluster length is 
\begin{equation}
    N\left(\theta\right) = \dfrac{\tan{\theta} \times \lambda }{55 \mum},
    \label{eq:tangus}
\end{equation}
where $N$ is the number of pixels forming the cluster and $\lambda$ is the active depth of the sensor. 
The measured cluster length for a given angle is distributed around the expected value.
A fit to the cluster length distribution is performed with a Gaussian distribution and clusters with length larger than one standard deviation from the fitted mean are rejected, removing about 40\% of the data.

Data sets have been acquired at four angles: 83, 85, 87, 89 degrees. 
The active depth of the sensor for nonirradiated and uniformly irradiated sensors is determined by performing a fit to the average cluster length using \eq\ref{eq:tangus}, where $\theta = \alpha + \epsilon$ with $\alpha$ fixed to the chosen angle and $\epsilon$ allowed to vary to account for a possible offset.
The angle offset is found to be of the order of $0.05$ degrees.

As illustrated in \fig\ref{fig:scheme}, the depth $d(\text{i})$, defined as the distance from the charge deposit to the pixel implant, can be parameterised by the hit position of the pixel $i$ within the cluster by inverting \eq\ref{eq:tangus}:
\begin{equation}
    d(i)= \frac{55\mum \times i}{\tan{\theta}}.
\end{equation}
Using this relationship, the time needed for the induced signal to  cross the threshold (\ttt) is investigated as a function of the depth.
A depth of $0 \mum$ corresponds to the pixel electrode side, while the full depth, $150 \mum$ or $200 \mum$ depending on the sensor, corresponds to the sensor back side.
The following plots are obtained with the sensor placed at a $85^{\circ}$ angle with respect to the beam, giving a depth step of $55\mum\times \cos(85^{\circ}) \approx 4.8\mum$, unless otherwise stated. 
The uncertainty on the measured depth due to the uncertainty on the angle offset and a possible missing hit at the beginning or end of the cluster is found to have a negligible impact on the charge and time-to-threshold distributions.

The charge collected is measured by performing a fit to the hit charge distribution at a given depth.
Two sources of systematic uncertainty on the MPV are considered, charge calibration and digitisation.
The systematic uncertainty on the charge calibration has two components: due to the imperfect knowledge of the injected charge and the statistical uncertainty from the test pulse procedure.
The former is assigned to be 4\% of the measured charge, according to Ref.~\cite{VicenteBarretoPinto:2134709}.
The latter is obtained by generating pseudoexperiments to evaluate how the correlated uncertainties of the calibration curve parameters affect the MPV and yields 30~\en for a nonirradiated sensor and 50~\en for a sensor irradiated at full fluence, where the uncertainty is larger in the irradiated case due to the smaller charge collected.
The digitisation uncertainty is assigned to account for the discrete values of ToT.
This uncertainty is estimated as 40~\en for hit charges higher than $\sim 2500$~\en, but rapidly increases for lower charges.
The effect of this systematic uncertainty on the MPV of the charge distribution is determined using pseudoexperiments and results in 20~\en and 50~\en for a nonirradiated sensor and for a sensor irradiated at full fluence, respectively.
\begin{figure}[!p]
  \centering
  {\includegraphics[width=0.49\textwidth]{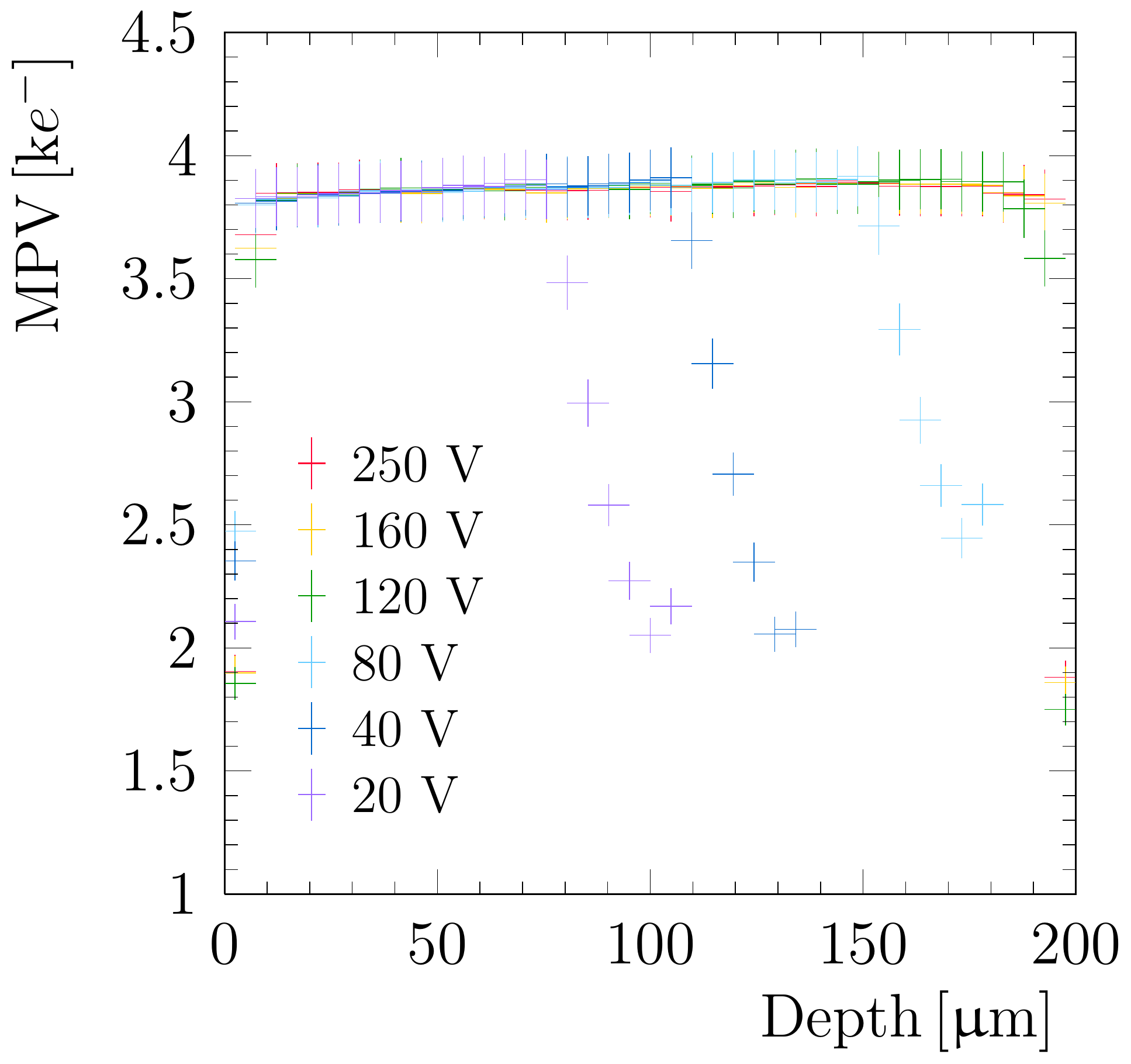}} \hspace{1mm}
  {\includegraphics[width=0.49\textwidth]{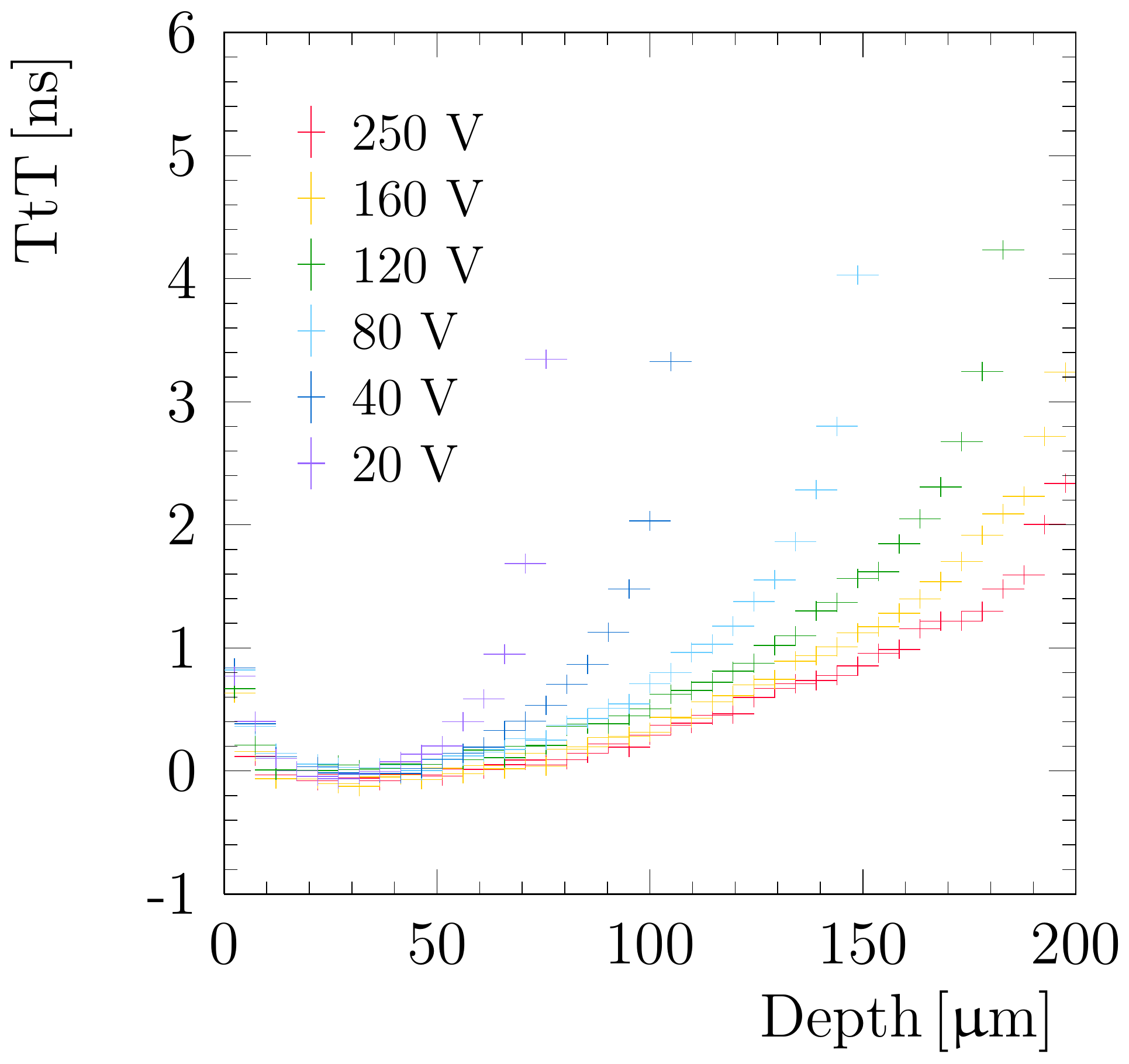}} \\
  {\includegraphics[width=0.49\textwidth]{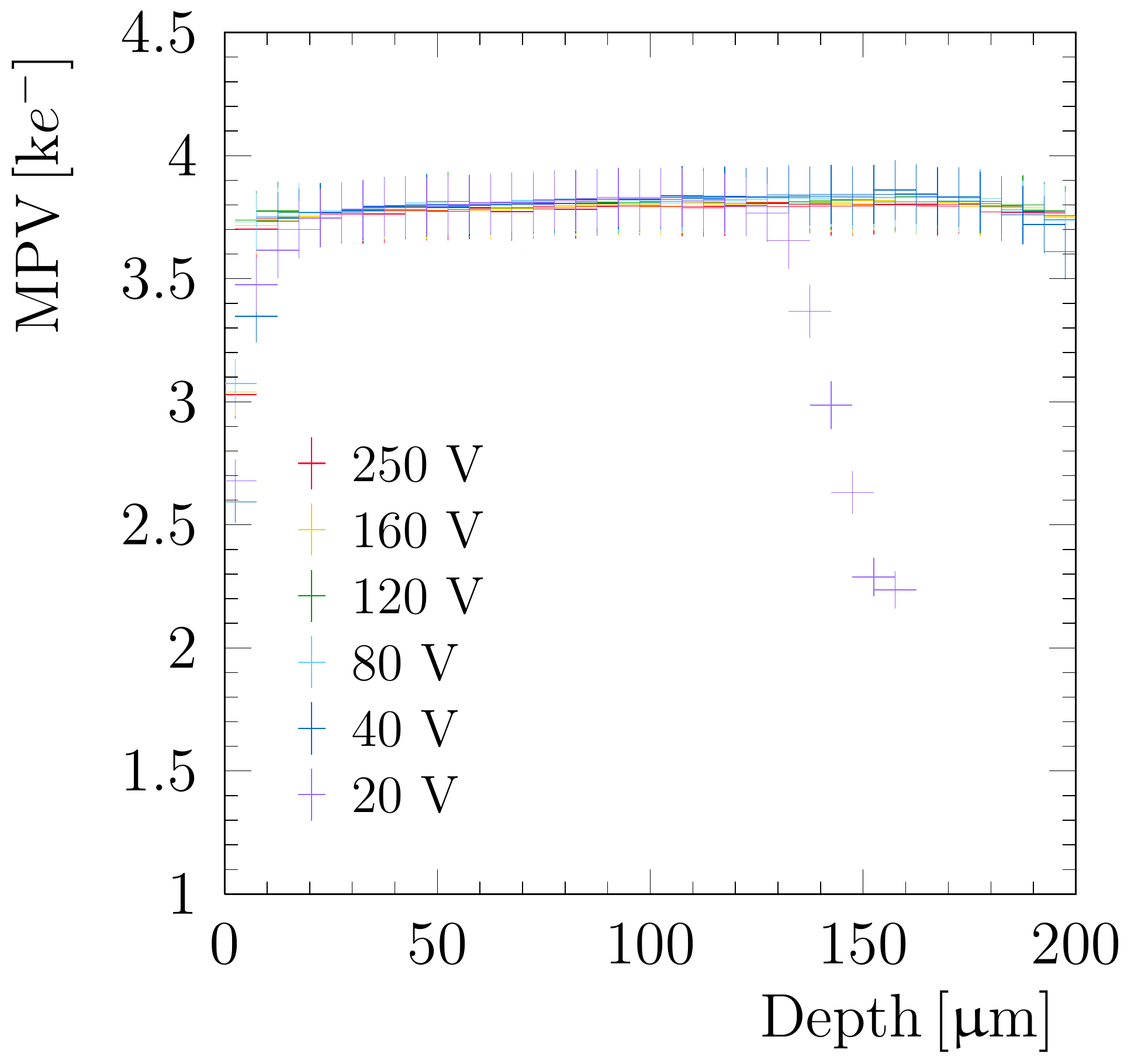}} \hspace{1mm}
  {\includegraphics[width=0.49\textwidth]{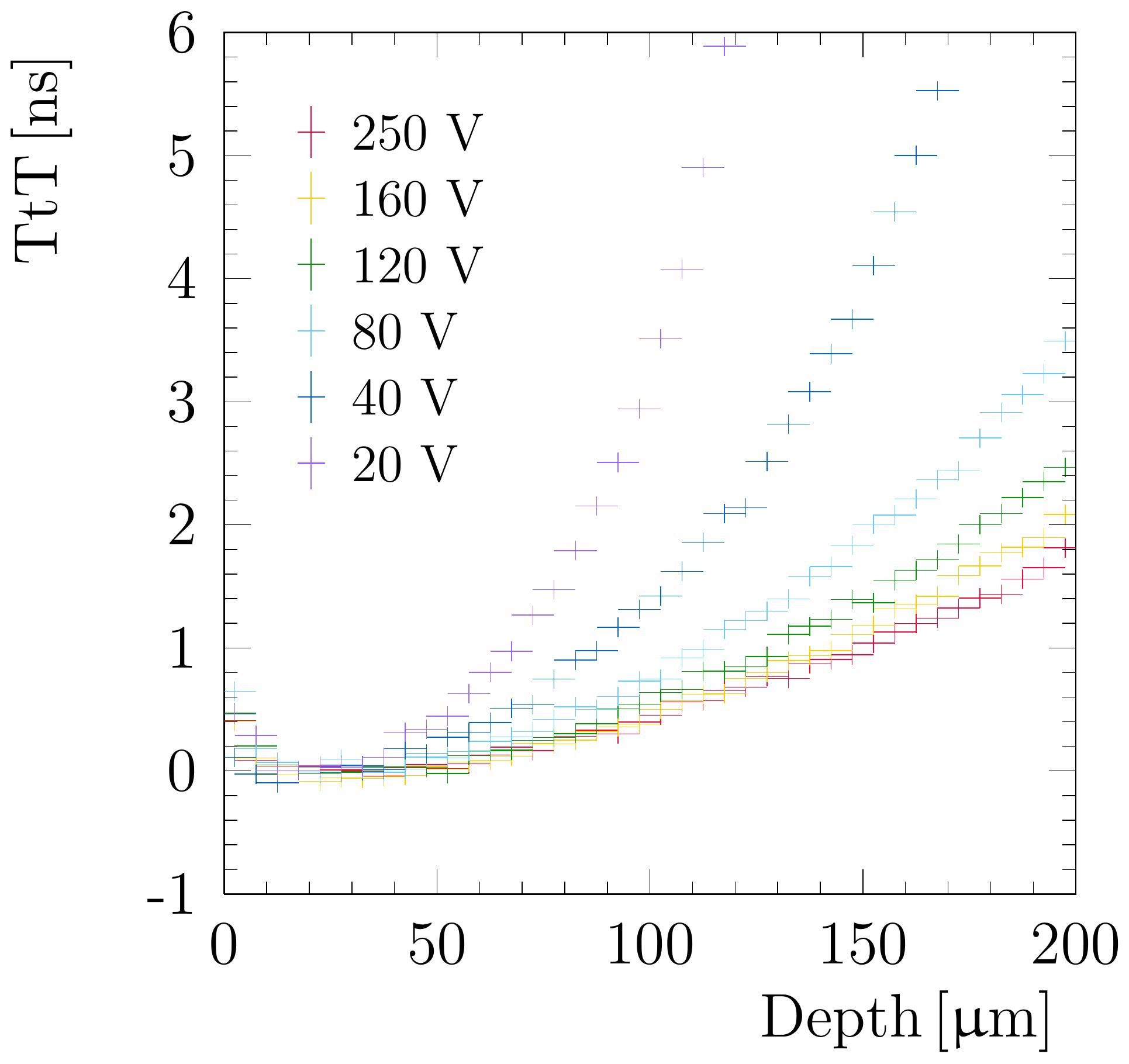}} \\
  {\includegraphics[width=0.49\textwidth]{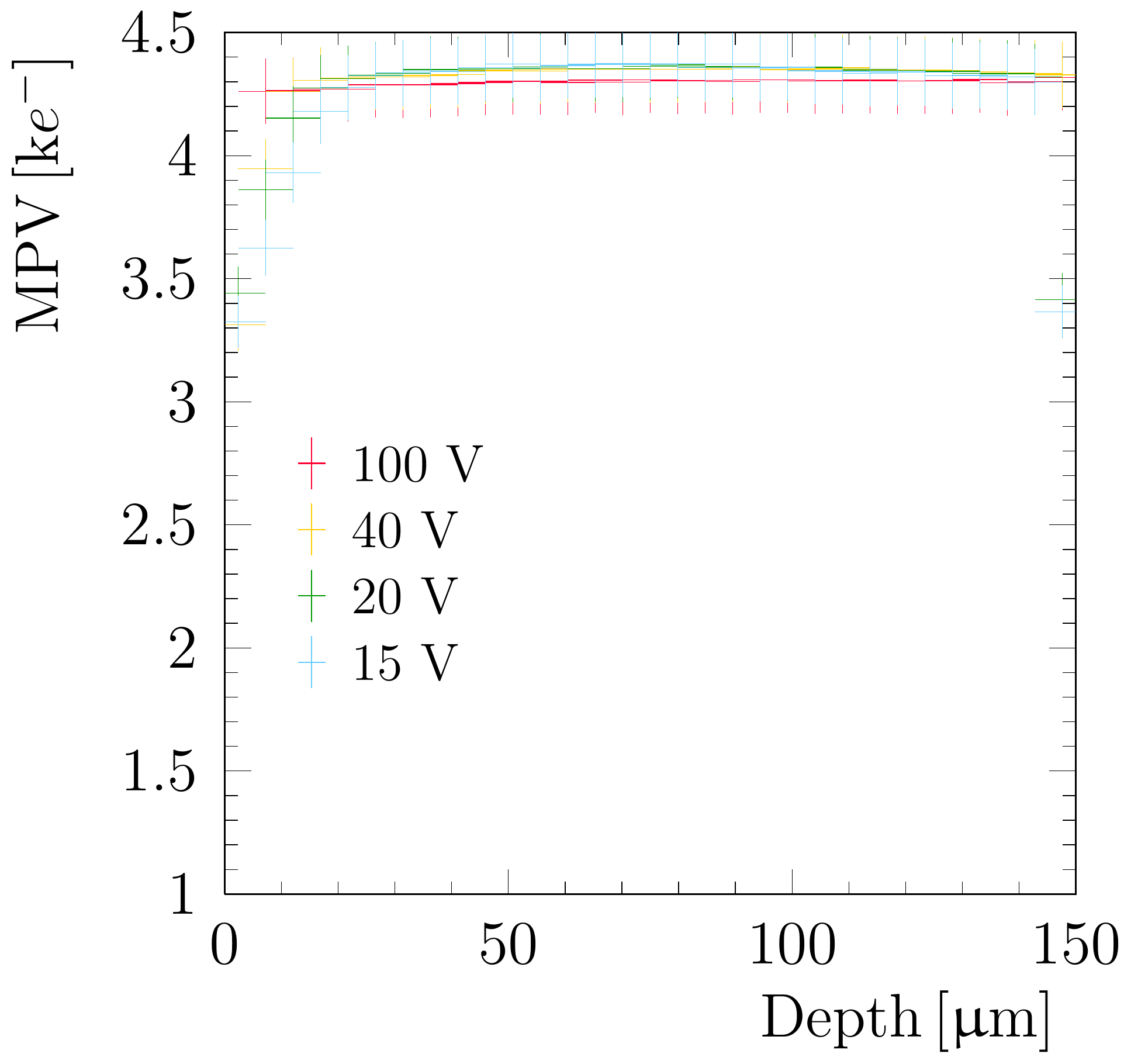}} \hspace{1mm}
  {\includegraphics[width=0.49\textwidth]{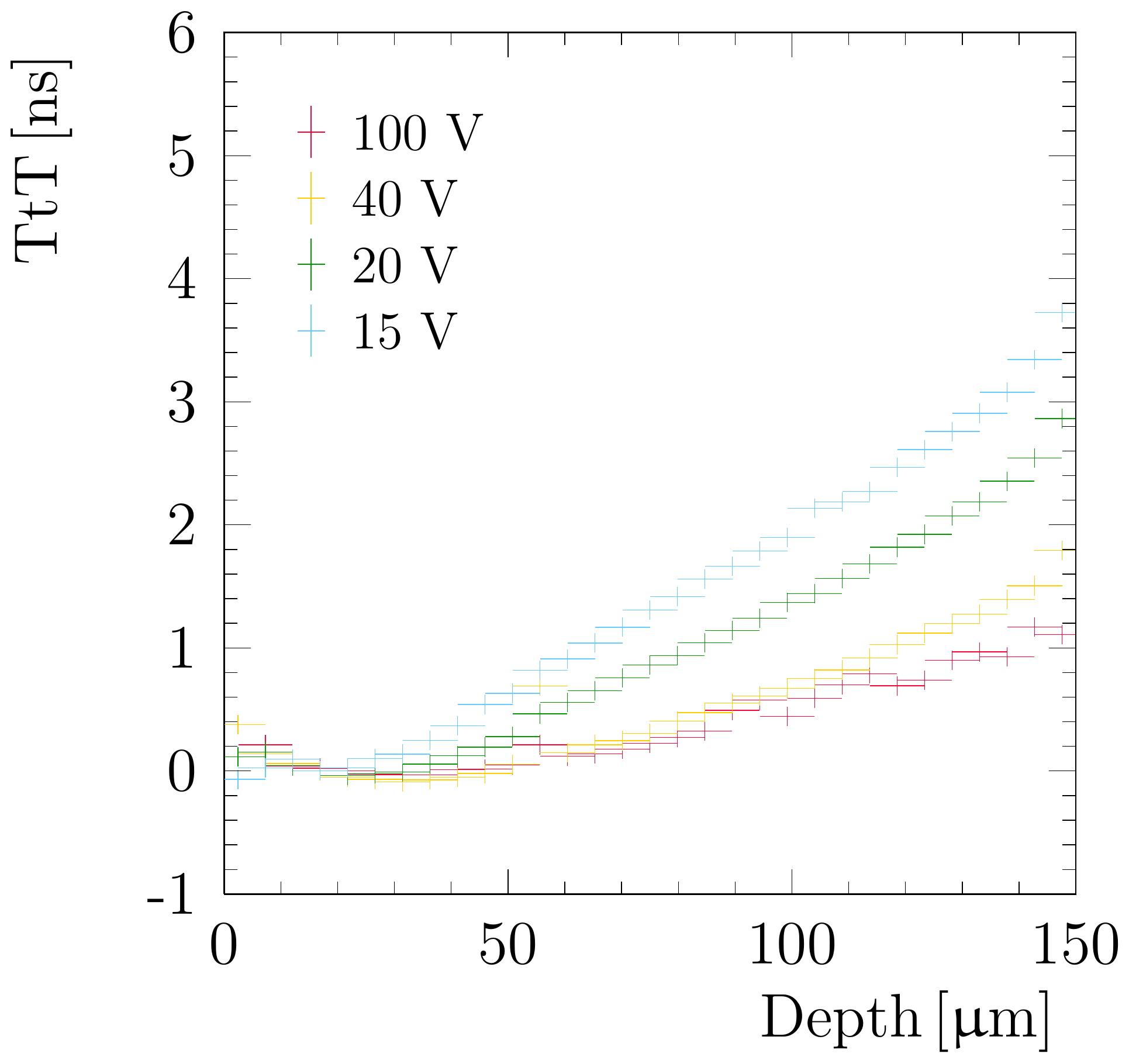}} 
  \caption{Charge collected (left) and \ttt (right) as a function of depth for a $200 \mum$ thick nonirradiated HPK \np sensor (S6, top), a $200 \mum$ thick nonirradiated Micron \np sensor (S23, middle) and a $150 \mum$ thick nonirradiated Micron \nn sensor (S34, bottom). A depth of 0~\mum corresponds to the pixel electrodes side while 200~\mum (top and middle) or 150~\mum (bottom)  corresponds to the backside.
  For the MPVs, the error bars indicate the systematic uncertainties on the measurements, while for the \ttt, the error bars indicate the statistical uncertainty.}
  \label{fig:MPVdepNonirr}
\end{figure}

\subsection{Nonirradiated sensors}
\label{subsec:grazingnonirr}

The results for \nonirr sensors are shown in \fig\ref{fig:MPVdepNonirr} for HPK \np (top), Micron \np (middle) and Micron \nn (bottom) sensors, in terms of charge collected (left) and \ttt (right) as a function of depth for different bias voltages.
It can be seen that the three families of sensors exhibit the same trend for both the charge collected and \ttt profiles.
For a \nonirr sensor and bias voltage above depletion, the MPV of the charge collected is constant and equal to the charge expected for the full thickness of the sensor.
The time needed to cross the threshold is less than $5 \ns$.
The depletion voltage is found to be 120~V and 40~V for HPK and Micron \np, respectively.
Below the depletion voltages the charge drops linearly, starting at the border between depleted and nondepleted volume up to a depth of about $20 \mum$ from the border.
This is an effect due to diffusion known as charge migration~\cite{CROITORU1985443,Tsopelas:2238509}. 
No hits are registered below 2000~\en, despite being significantly far from the threshold, due to a time requirement of $100\ns$ imposed in the clustering process.
The increase in \ttt with depth can be mainly attributed to the nonuniformity of the weighting field, which increases towards the pixel electrodes, and hence most of the signal is induced while drifting near the electrodes.
Since the sensor is nonirradiated and the collected charge is higher than 3000~\en, the timewalk has a negligible effect. 
Partially depleted sensors have an additional contribution to the time-to-threshold from charges migrating from the nondepleted region due to diffusion. 

\begin{figure}[b!]
  \centering
  {\includegraphics[width=0.49\textwidth]{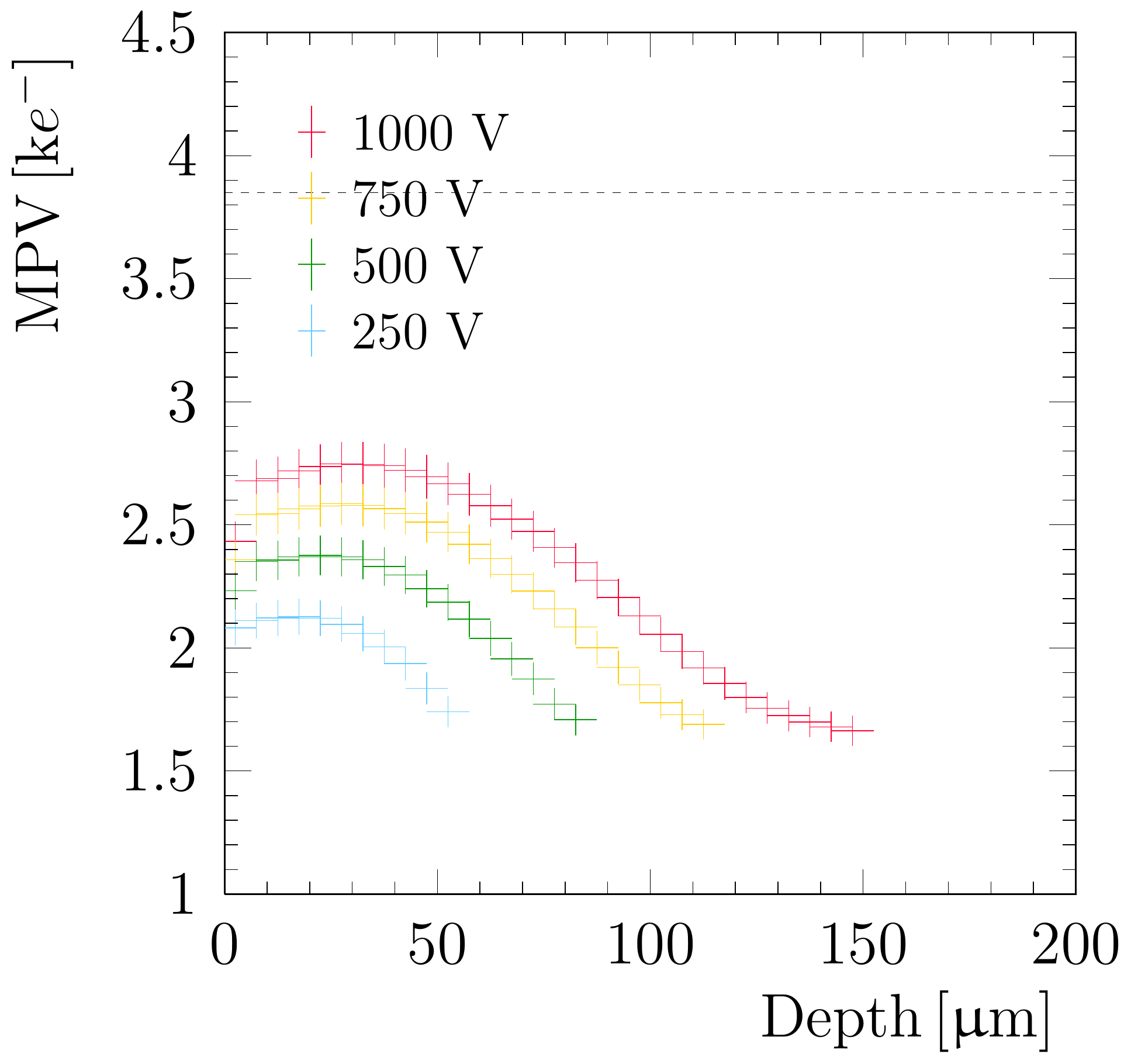}} \hspace{1mm}
  {\includegraphics[width=0.49\textwidth]{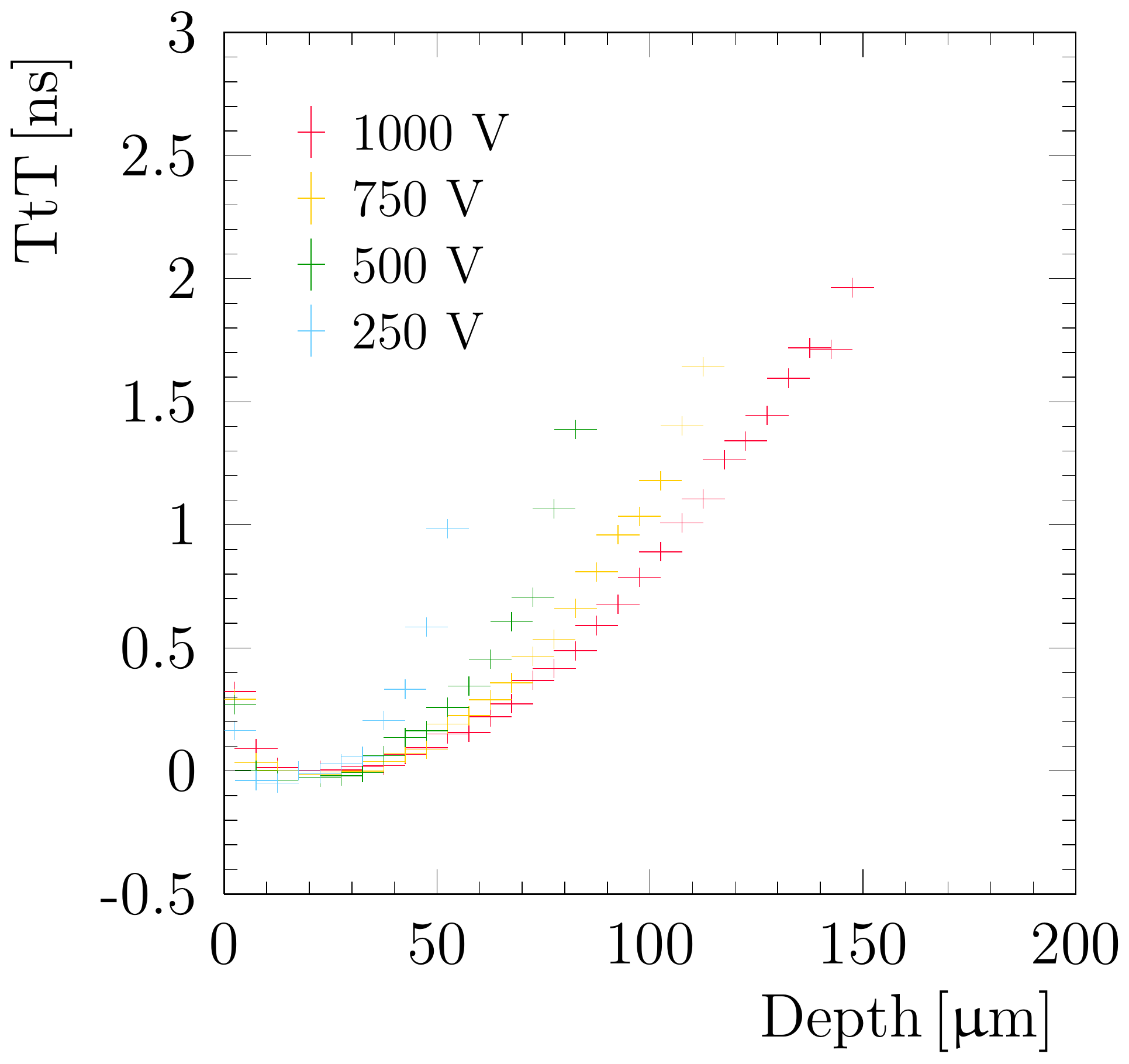}} \\
  \caption{Charge collected (left) and \ttt (right) as a function of depth for a $200 \mum$ thick  HPK \np sensor (S22).
   The dashed line indicates the charge collected by a nonirradiated sensor of the same type. The sensor is uniformly irradiated to \maxfluence.
  For the MPVs, the error bars indicate the systematic uncertainties on the measurements, while for the TtT, the error bars indicate the statistical uncertainty.} 
  \label{fig:MPVdepIrr}
\end{figure}
\begin{figure}[b!]
    \centerline{\includegraphics[width=0.49\textwidth]{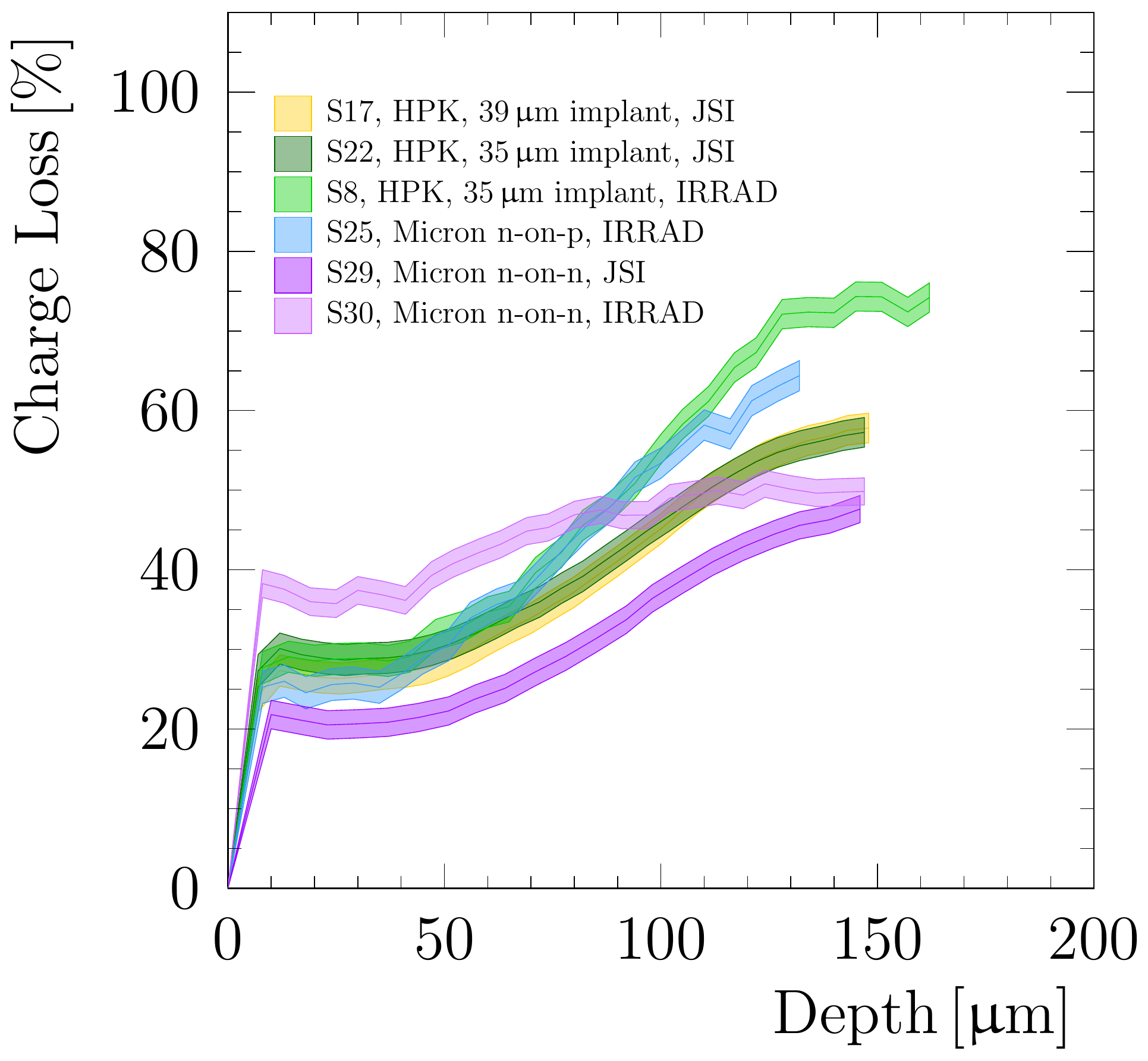}} 
    \caption{Percentage of charge loss as a function of depth for a HPK \np and a Micron \np sensors and for different radiation types.}
  \label{fig:chargeloss}
\end{figure}

\subsection{Sensors irradiated to full fluence}
\label{subsec:grazingunifirr}

\begin{figure}[b!]
  \centering
  {\includegraphics[width=0.49\textwidth]{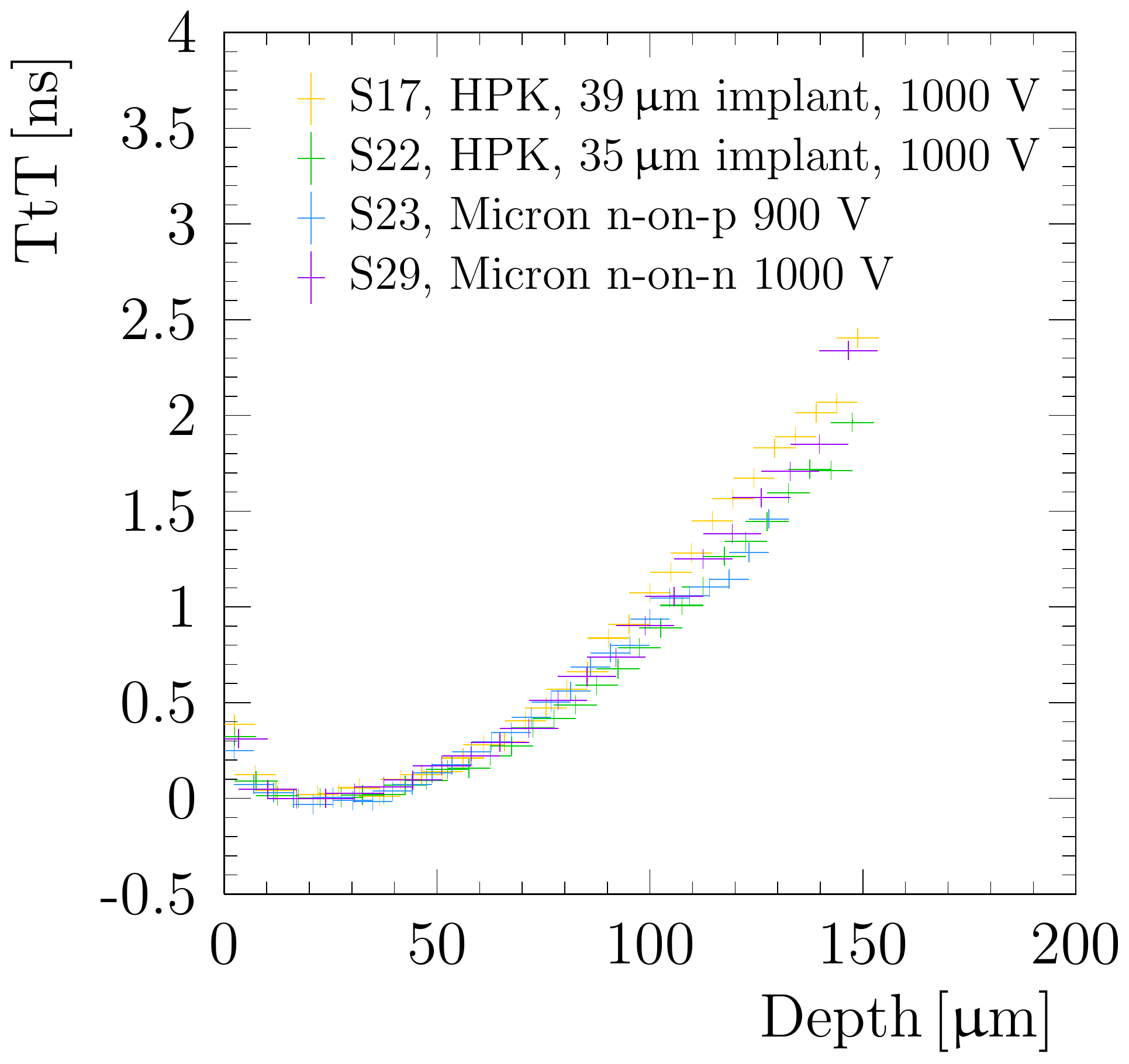}} \hspace{1mm}
  {\includegraphics[width=0.49\textwidth]{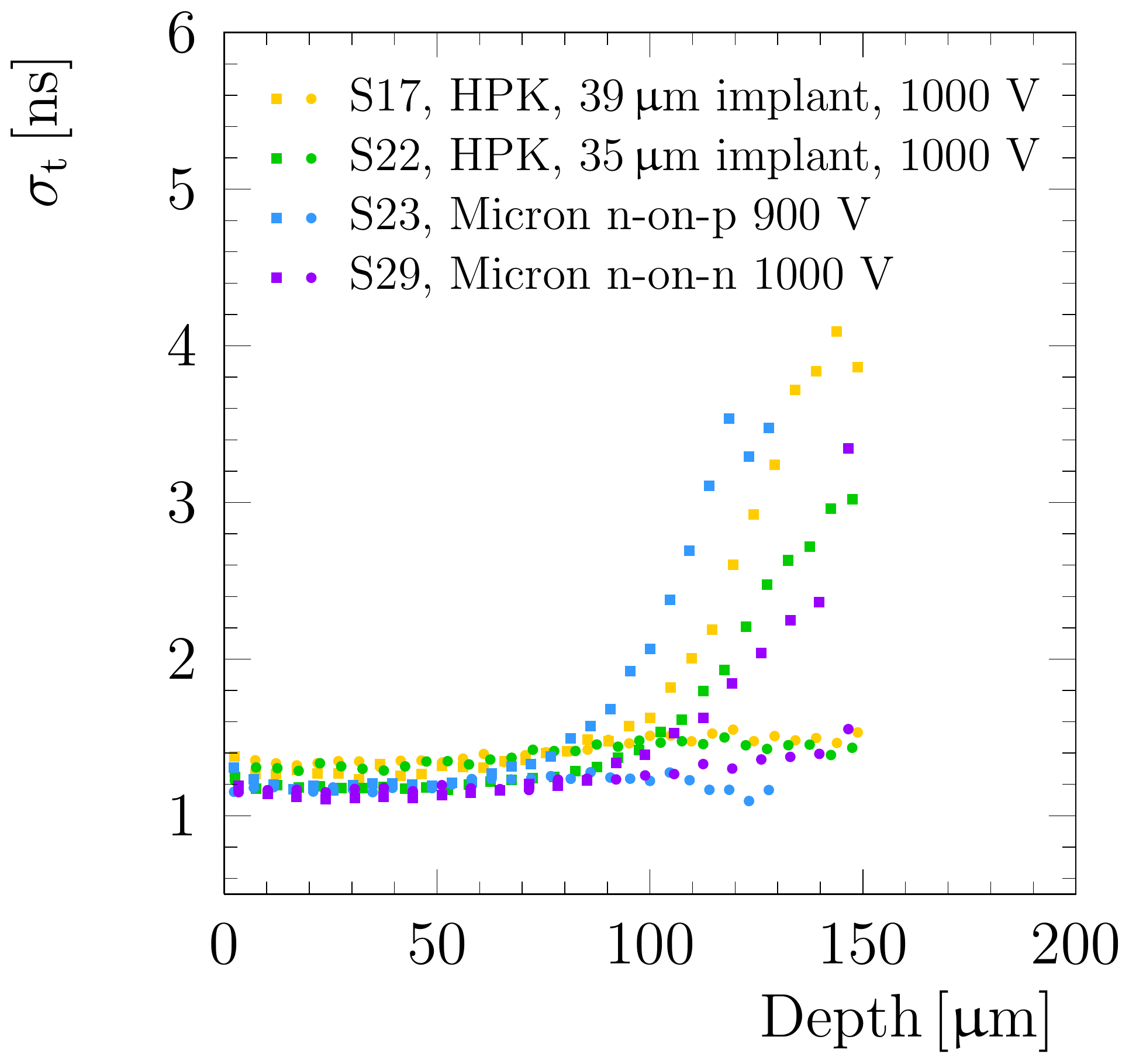}} \\
  \caption{Time-to-threshold (left) and resolution (right) as a function of depth for uniformly irradiated sensors.
  The circle and square markers indicate the left and right-sided resolution, respectively.
 } 
  \label{fig:TtTIrrcomp}
\end{figure}
The trends of the charge collected and \ttt as a function of depth for uniformly irradiated sensors, as illustrated in \fig\ref{fig:MPVdepIrr} for HPK \np, are quite different from the behaviour observed for \nonirr devices. 
Firstly, there is an overall signal reduction due to charge trapping, which is not fully recovered by an increase in bias voltage. 
Not all the charge liberated at a given depth is collected and this decreases with distance from the electrodes because charge needs to travel over a longer distance and hence has a larger probability to be trapped~\cite{Moll:2640820}.
The charge at each bias voltage shows a slight decrease close to the electrodes. 
This is due to two different effects: the lower field between the neighbouring pixel implants and hole trapping, since for charges liberated close to the electrodes the current is mainly induced by the motion of holes.
Secondly, most of the charge from the nondepleted volume recombines before it could be collected.
This can be attributed to charge trapping and slow drift in combination with the integration time of the front-end. 
The time for the integration of the signal is limited, hence the discharge can start while still in the process of integrating; this is especially relevant for a small amount of charge.
A possible effect of a doubly peaked electric field according to the double junction model for highly irradiated sensors~\cite{Moll:2640820,Eremin:2002wq} is not observed.
This can be attributed to a combination of the small amount of charge and the low weighting field at the backside of the sensor.

The percentage of charge loss due to irradiation per depth is illustrated in \fig\ref{fig:chargeloss} for the different types of sensors and different types of irradiation, where for the sensors nonuniformly irradiated at IRRAD the region with average fluence of $7.6 \times \fluence $ is selected. 
The percentage of charge lost varies with depth between about 25\% close to the electrodes and 60\% at the border of the active region. 
These values are compatible with what is reported in literature for similar fluences~\cite{Ducourthial:2018mgq}.
The charge loss increases with depth up to the active volume of the sensor after irradiation, while nothing is collected from the nondepleted volume. 
Only Micron \nn sensors ($150\mum$ thick) reach full depletion.
A different behaviour is observed between proton and neutron irradiation, with a steeper dependence on depth in the case of protons. 

The \ttt after timewalk correction increases with depth, up to $2 \ns$, independent of the voltage applied.
The different sensors show the same trend in \ttt as a function of depth, as can be seen in \fig\ref{fig:TtTIrrcomp} (left).
The resolution as a function of depth is shown in \fig\ref{fig:TtTIrrcomp} (right) and is found to be constant up to around $90\mum$ depth. 
At this point, the right-sided resolution increases due to residual timewalk. 
Timewalk corrections are large for these assemblies and therefore the results prior to correction are described in detail in Appendix~\ref{app:timewalk}.

\subsection{Nonuniformly irradiated sensors}
\label{subsec:grazingnonunifirr}

The assemblies presented in this section were irradiated nonuniformly, following the shape of the illumination by the proton beam as shown in \fig\ref{fig:fluence}. 
All the assemblies have been tested without additional controlled annealing, with the exception of the Micron \nn sensor that underwent controlled annealing for 80 minutes at $60 \degrees$C.

\begin{figure}[b!]
  \centering
  {\includegraphics[width=0.49\textwidth]{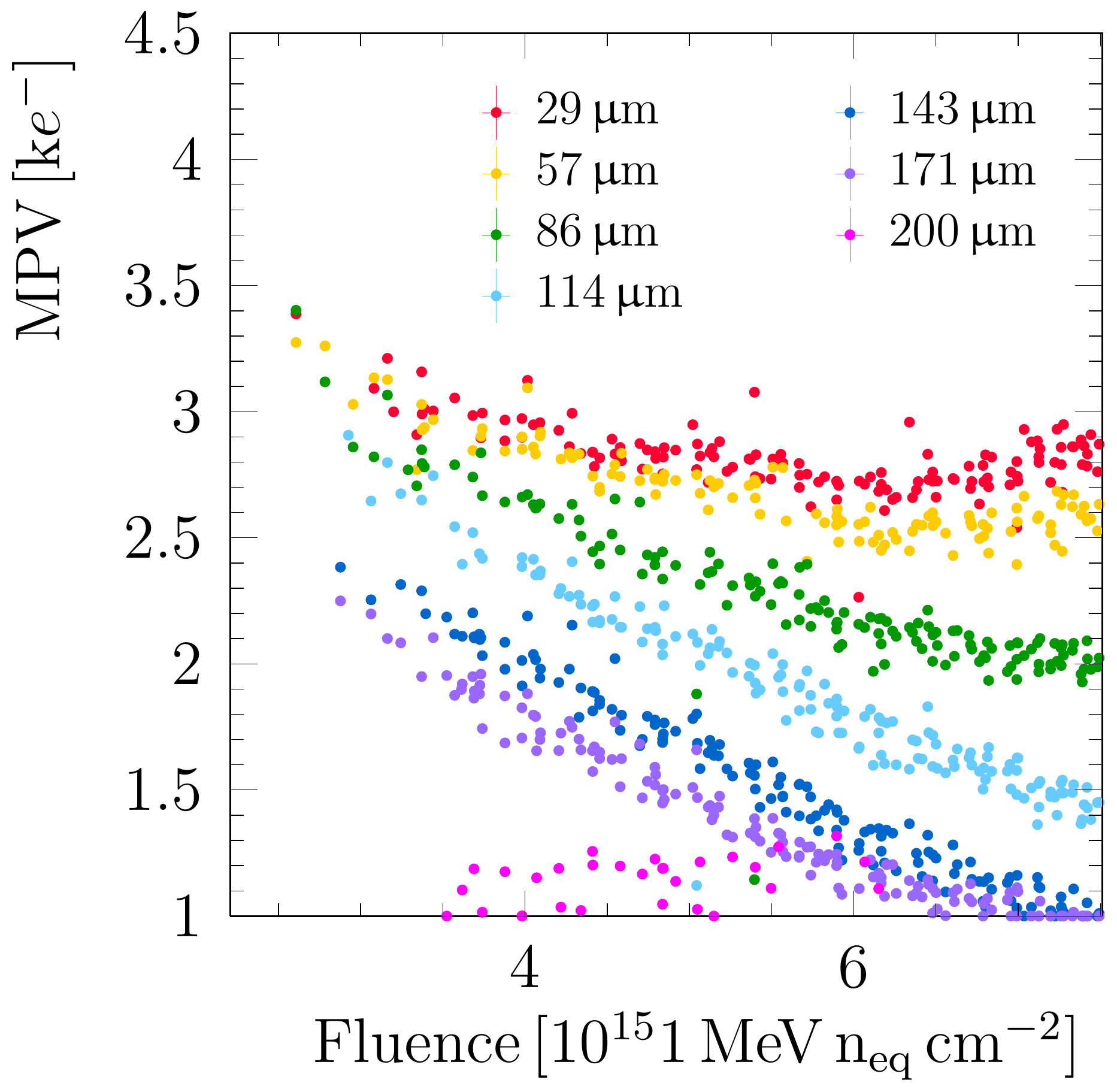}} \hspace{1mm}
  {\includegraphics[width=0.49\textwidth]{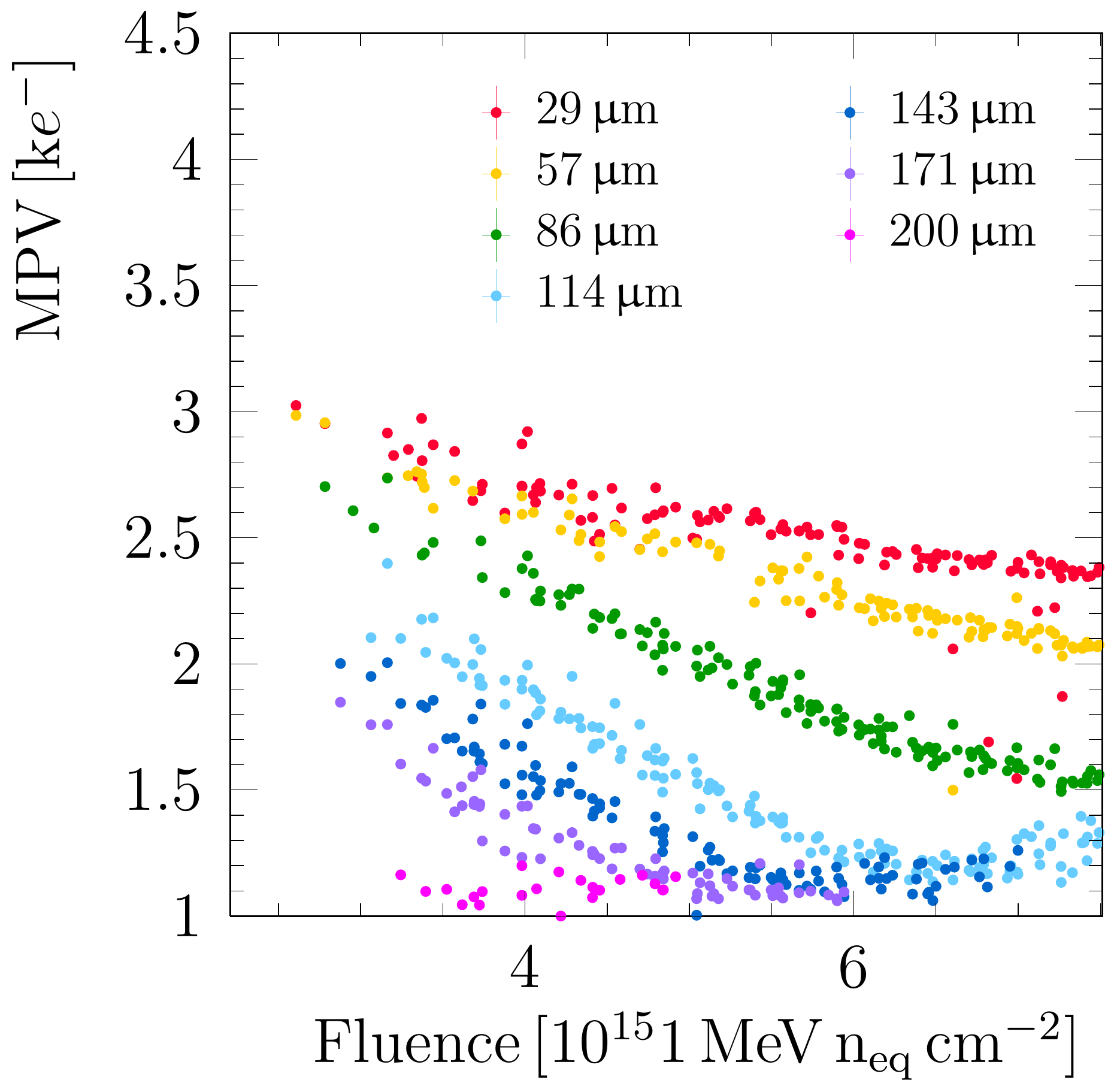}} \\
  {\includegraphics[width=0.49\textwidth]{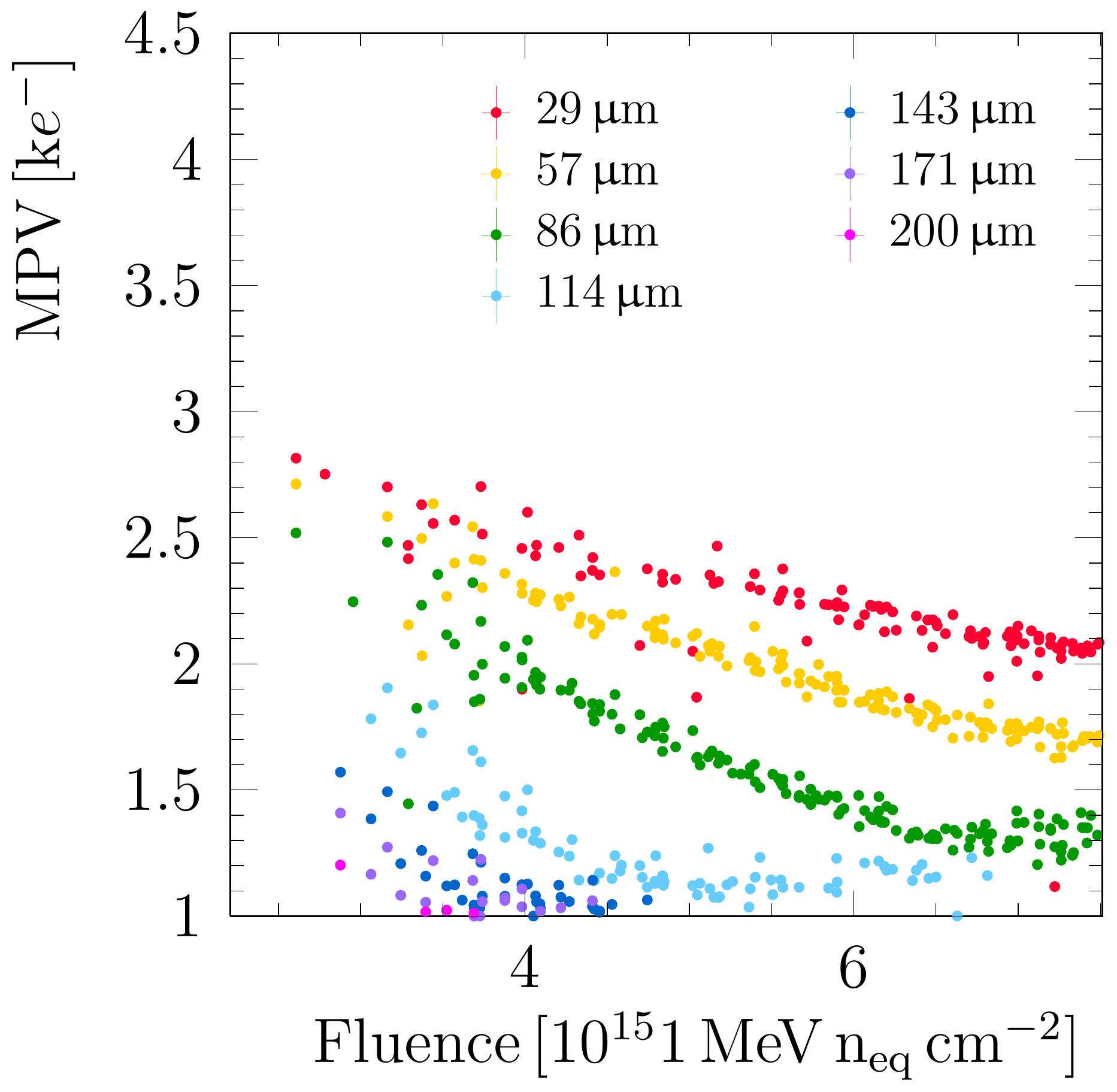}} \hspace{1mm}
  {\includegraphics[width=0.49\textwidth]{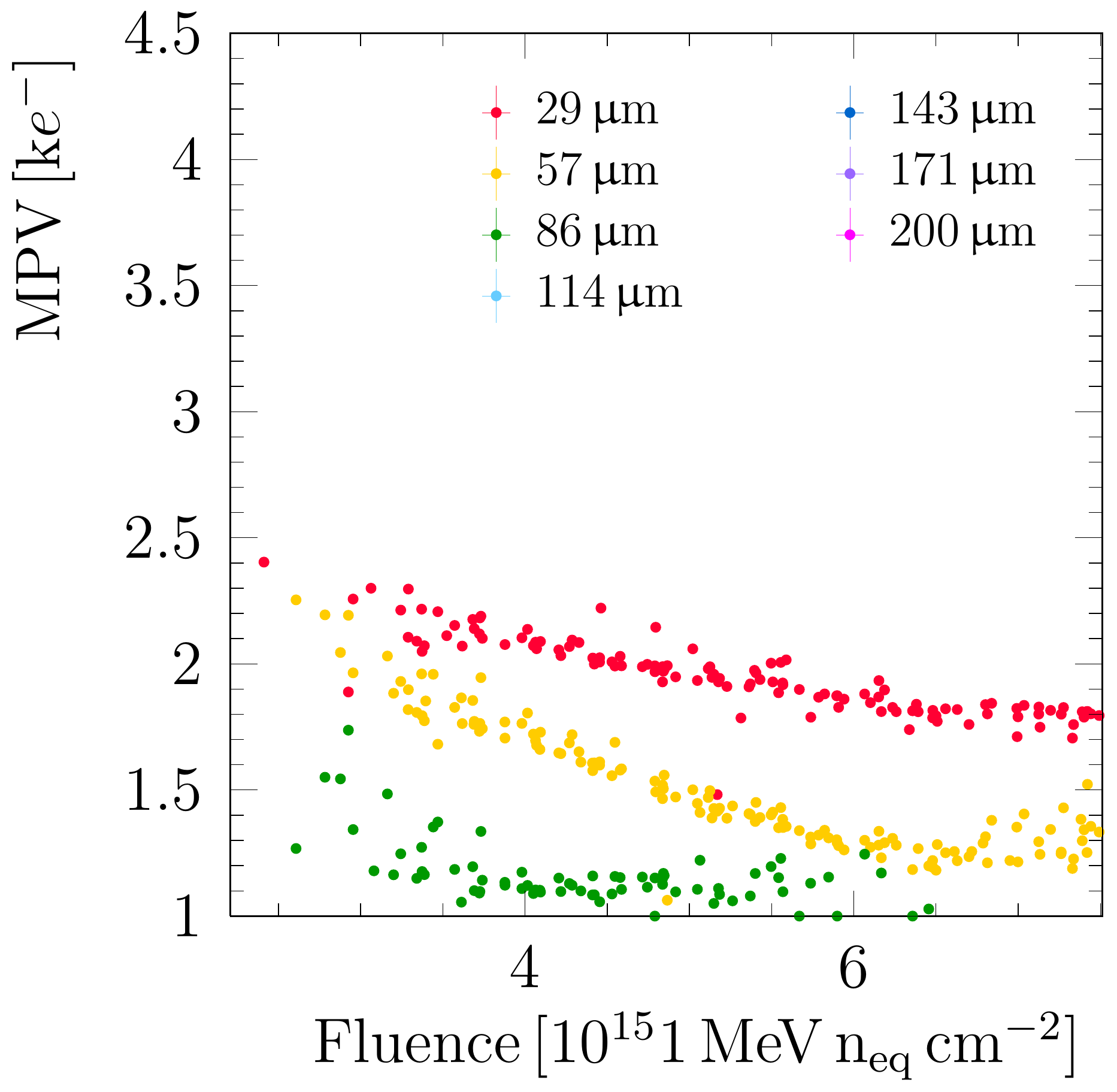}} \\ 
  \caption{Charge collected as a function of fluence from different depths for a $200 \mum$ HPK \np sensor (S8) operated at 1000 V (top left), at 750 V (top right),  at 500 V (bottom left) and at 250 V (bottom right).}
  \label{fig:MPVvsFluenceS8}
\end{figure}

\begin{figure}[t!]
  \centering
  {\includegraphics[width=0.49\textwidth]{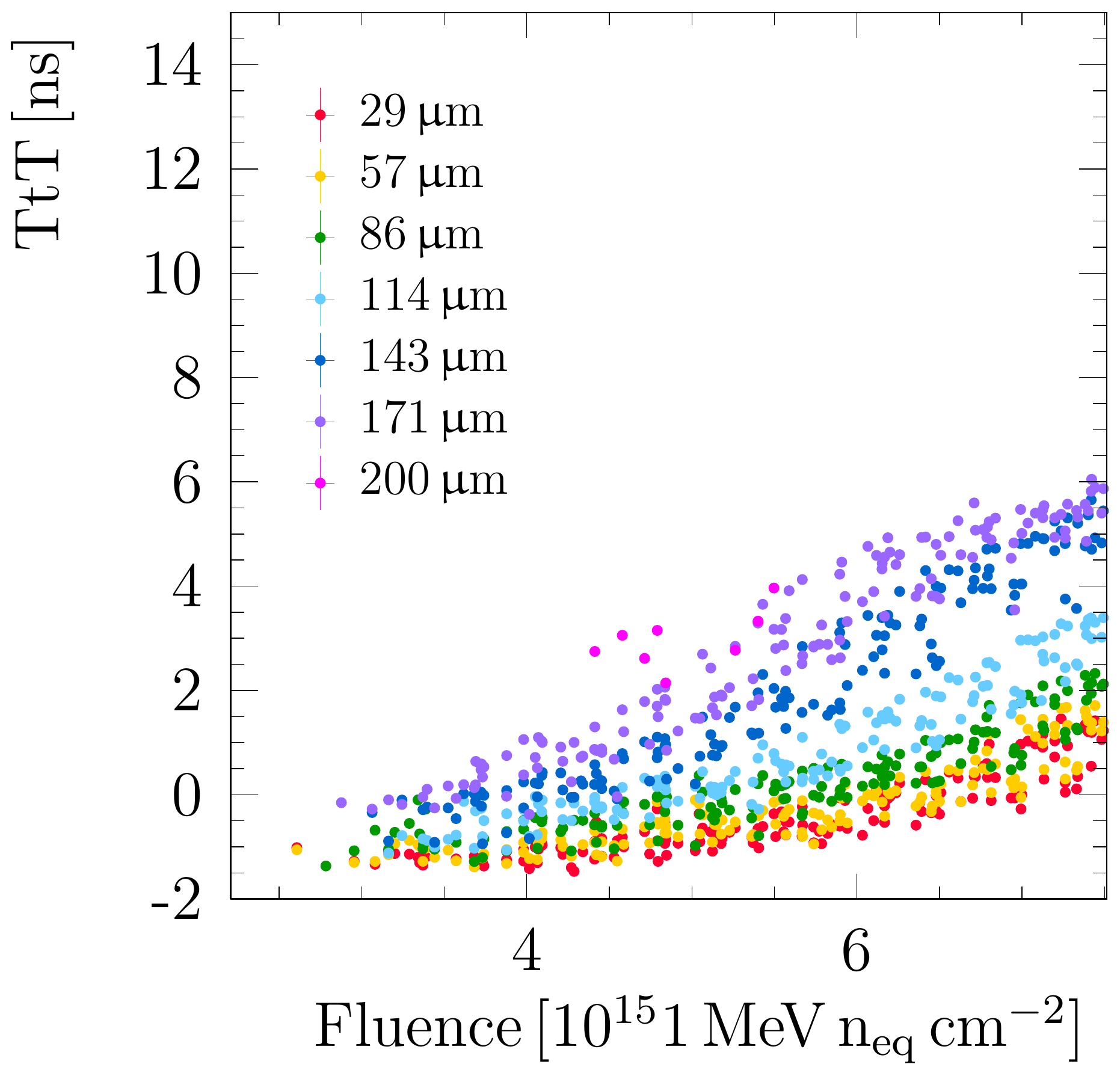}} \hspace{1mm}
  {\includegraphics[width=0.49\textwidth]{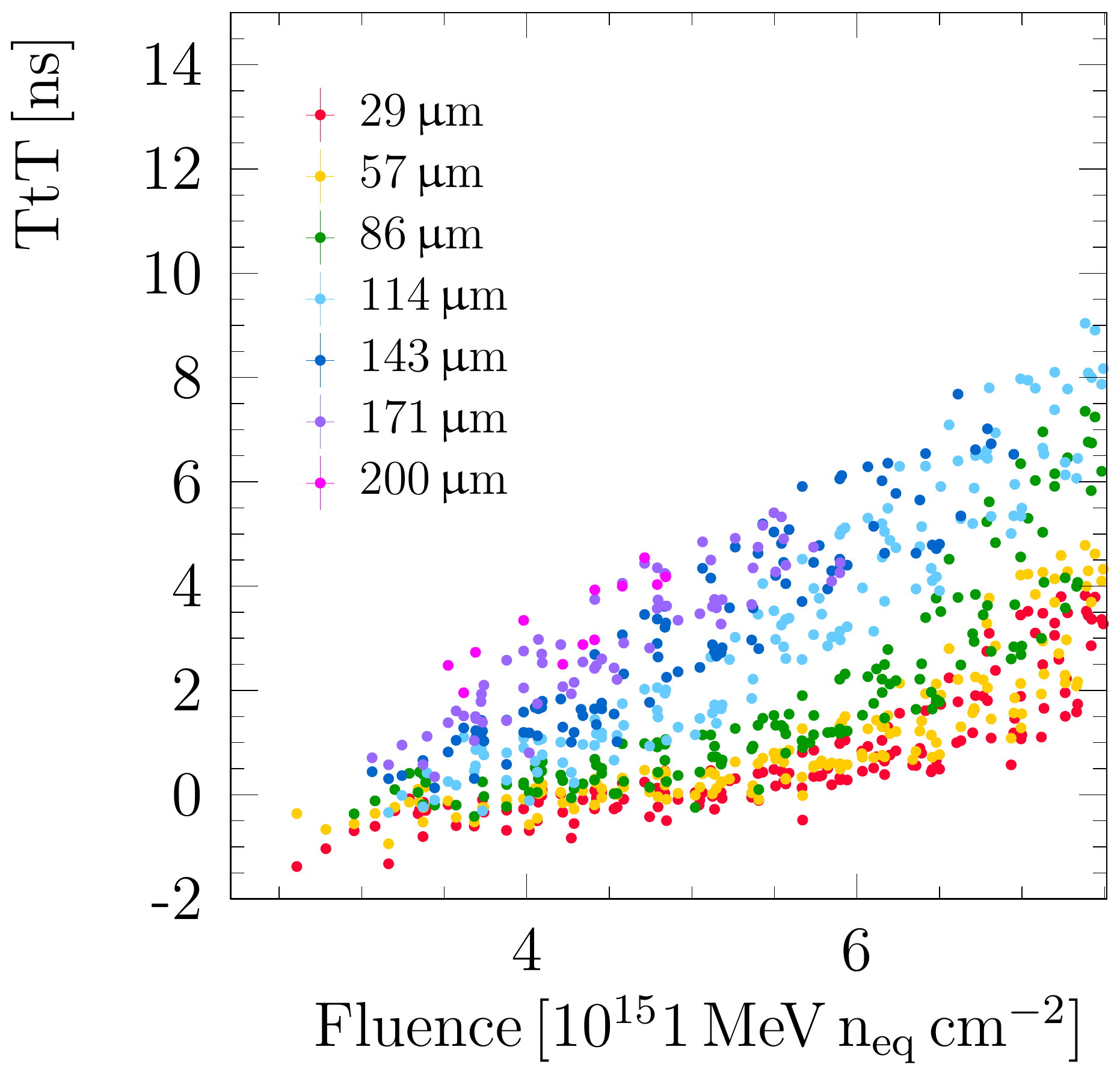}} \\ 
  {\includegraphics[width=0.49\textwidth]{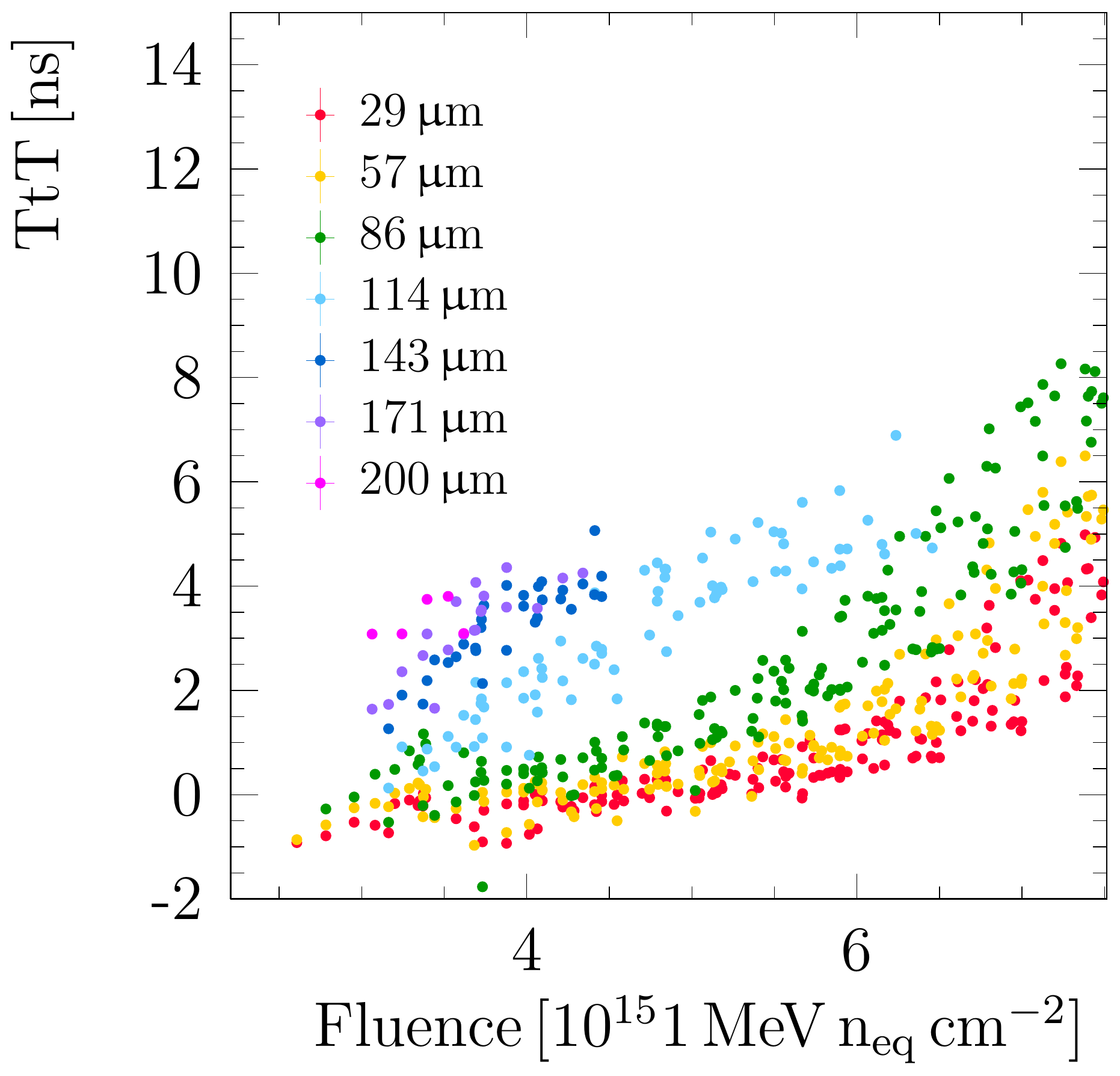}} \hspace{1mm}
  {\includegraphics[width=0.49\textwidth]{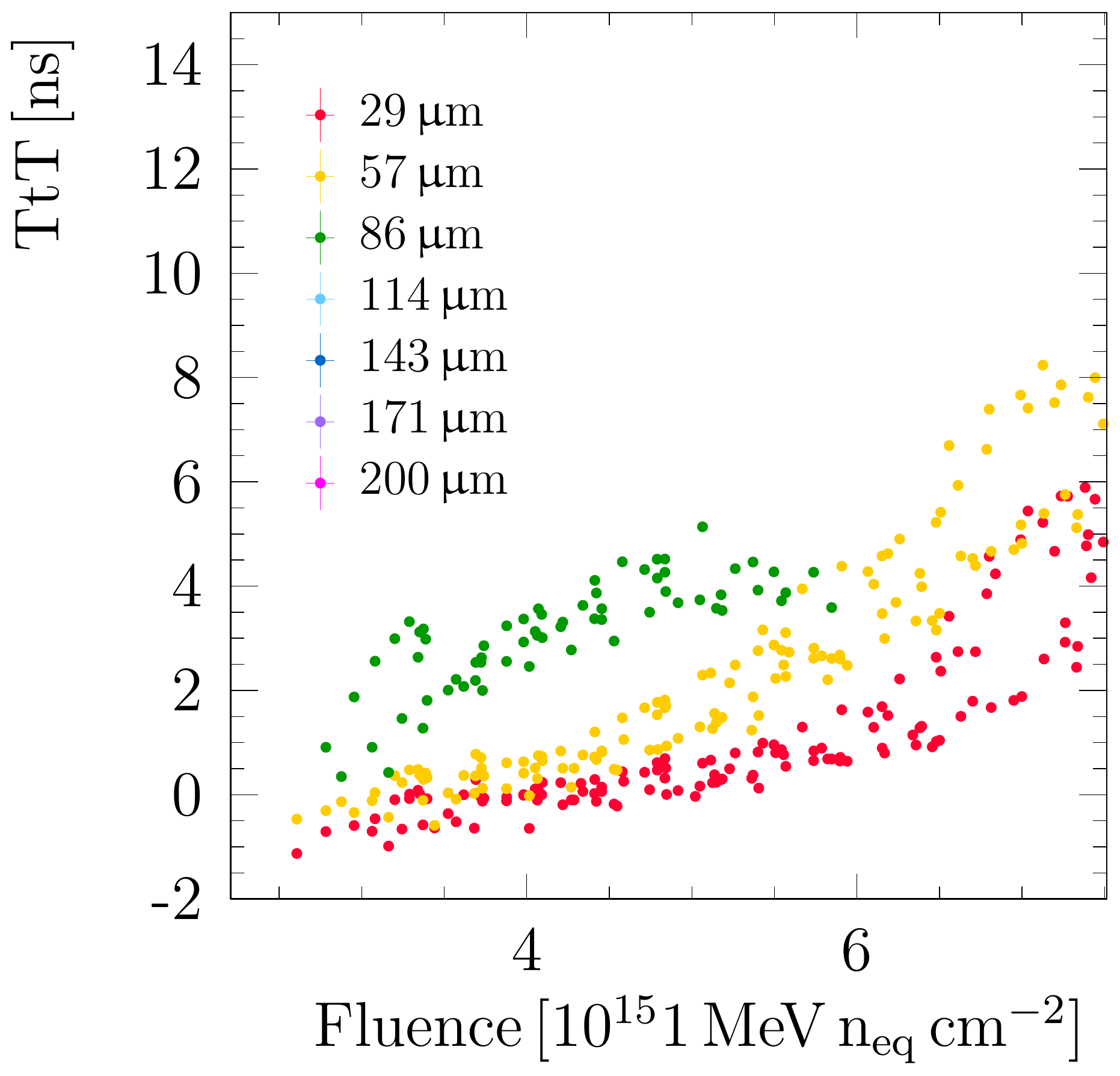}} \\ 
  \caption{Time-to-threshold as a function of fluence from different depths for a $200 \mum$ HPK \np sensor (S8) operated at 1000 V (top left), at 750 V (top right), at 500 V (bottom left) and at 250 V (bottom right).}
  \label{fig:TtTvsFluenceS8}
\end{figure}

The charge collection and \ttt for charges liberated at different depths in the bulk of the sensors are studied as a function of fluence at different operation voltages in \fig\ref{fig:MPVvsFluenceS8} and \fig\ref{fig:TtTvsFluenceS8} for a HPK \np sensor. 
The charge is collected only up to $\sim 90 \mum$ depth from the electrodes for low voltages, around 250~V, and decreases as a function of fluence.
Increasing the bias voltage, the charge is collected from deeper in the sensor, 
up to $\sim 170 \mum$ depth at 1000~V.
At the highest voltage tested, the spread in \ttt with depth is about $1\ns$ at low fluences and about $6 \ns$ at \maxfluence, with charge collected more slowly from deeper in the sensor.
At a given depth and fluence charge is collected more slowly as the voltage is decreased.
Larger data sets would be needed to determine any changes in the resolution of the sensor.

At depths smaller than \SI{80}{\micro\meter}, there is a sizeable deviation from the trend for the highest fluence region.
This effect, observed only at 1000~V bias, is recognised as the so-called charge multiplication or avalanche effect \cite{MIKUZ2011S50,CASSE2011S56,Lange:2010zz}. 
The charge multiplication happens close to the pixel electrodes where the electric field is highest.
If charges are liberated in the high field volume, close to the electrodes, they have higher chance to undergo multiplication. 
In contrast, most of the charges liberated deeper in the bulk will experience trapping before reaching the high field region.
A small rise in the apparent MPV is also observed for charges close to the threshold, at a depth that varies as a function of applied voltage. This rise can be attributed to a threshold effect, an investigation into which is documented in Appendix~\ref{app:threshold}.
Avalanche multiplication is observed in both Hamamatsu \np and Micron \np sensors at 1000 V, while the Micron \nn sensor does not present any observable effect even at the highest voltage tested.
The charge multiplication effect is not observed for uniformly neutron irradiated sensors, even when operated at 1000 V.
This can be attributed to the different nature of irradiation leading to different damage in the silicon~\cite{junkes2011status}.

\begin{figure}[t!]
  \centering
  {\includegraphics[width=0.49\textwidth]{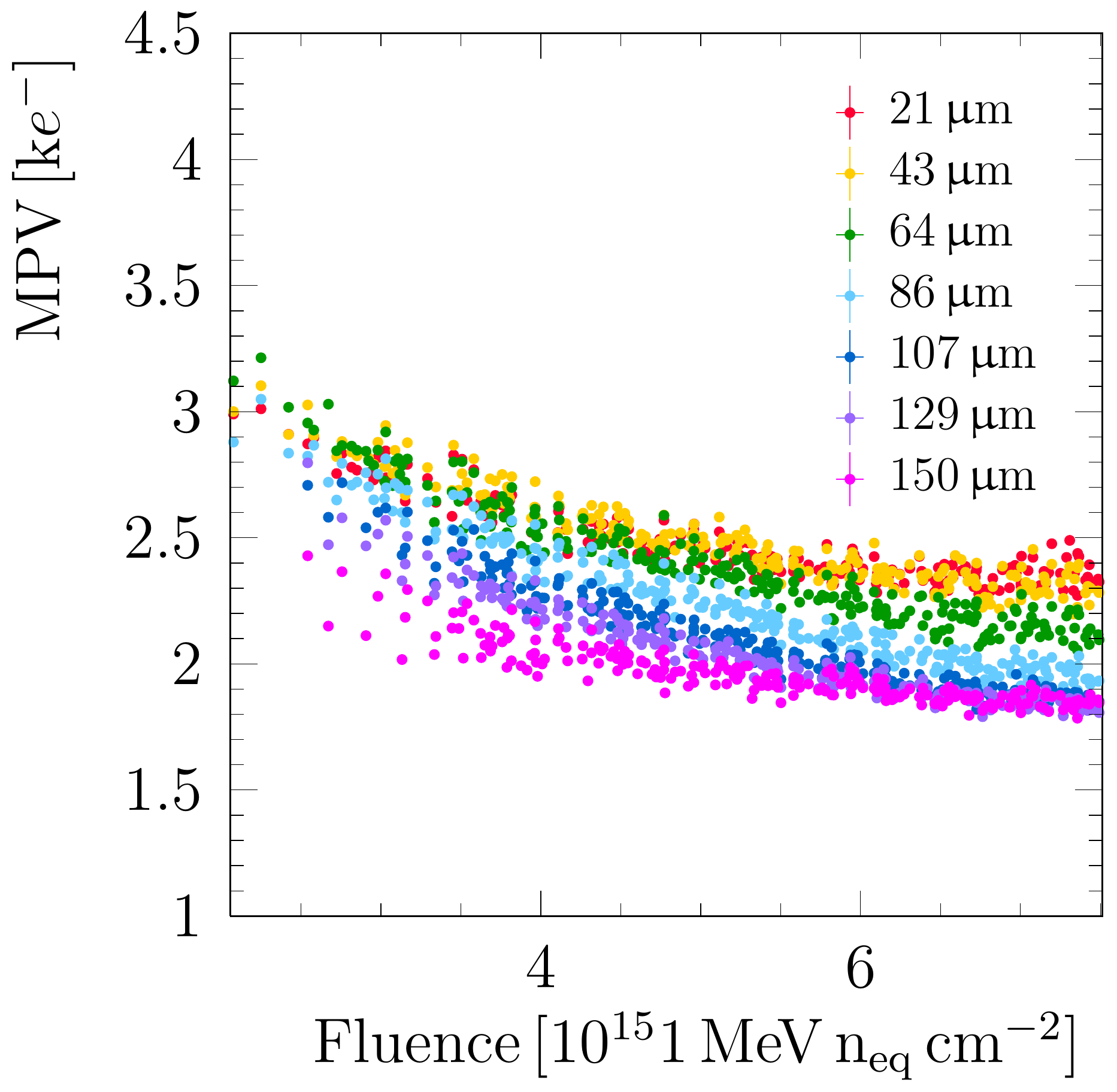}}\hspace{1mm}
  {\includegraphics[width=0.49\textwidth]{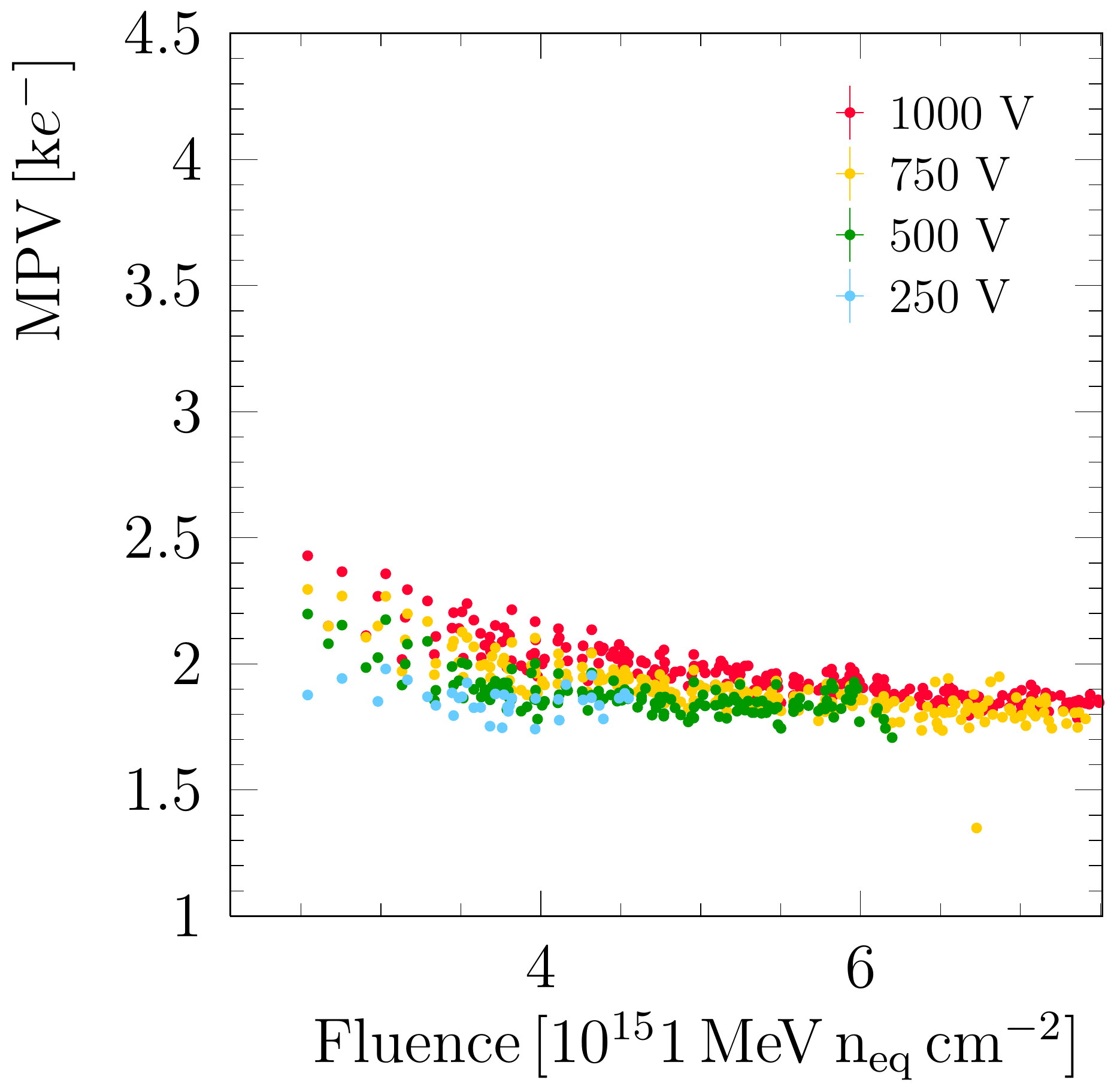}}
  \caption{Charge collected as a function of fluence from different depths for a sensor operated at 1000~V (left) and charge collected as a function of fluence for charges liberated at $150 \mum$ depths at different operation voltages (right). The sensor is a $150 \mum$ Micron \nn sensor (S30).}
  \label{fig:MPVvsFluenceS30}
\end{figure}

In addition to the absence of the charge multiplication effect, \nn sensors present a lower degradation as a function of the fluence, as can be seen from \fig\ref{fig:MPVvsFluenceS30}.
The charge collected at a depth of $150 \mum$ is examined as a function of fluence for different operation voltages in \fig\ref{fig:MPVvsFluenceS30} (right).
It can be seen that the charge collected never falls below the threshold value of 1000~\en and thus is collected up to the full thickness even at 400V, showing that Micron \nn has a larger active depth compared to the other types of sensors.

\FloatBarrier
%%%%%%%%%%%%%%%%%%%%%%%%%%%%%%%%%%%%%%%%%%%%%%%%%%%%%%%%%%%%%%%%%%%%%%%%%%%%%%%%%%%%%%%%
\section{Conclusions}
\label{sec:conclusion}

The timing properties of a range of prototype sensors are presented in this paper. 
Several assemblies are studied perpendicular to the incident beam in order to investigate the sensor resolution and \ttt as a function of bias voltage and fluence. 
Before irradiation the temporal resolution saturates at about $0.8\ns$ slightly before reaching full depletion. 
After uniform irradiation up to a fluence of \maxfluence, the temporal resolution does not saturate up to at least 1000~V. 
Assemblies irradiated with a non-uniform irradiation profile show that the temporal resolution degrades with increasing fluence. 
It is also observed that with the onset of charge multiplication, the temporal resolution does not improve. 

The grazing angle technique proves to be a powerful method to study charge collection and \ttt properties of charges generated at different depths in the bulk of the sensors. 
For nonirradiated sensors the most probable value of collected charge is constant as a function of depth once full depletion is reached and higher than 3500~\en.
The \ttt is therefore barely affected by timewalk.
The \ttt is extended by the time needed for the charge to migrate from the nondepleted volume.
For sensors uniformly irradiated to the full fluence it is observed that most of the charge collected originates close to the pixel electrode.
Due to radiation damage there is  a reduction of the charge collected, leading to an increase in timewalk. Otherwise, the temporal collection properties of the sensor are only marginally affected by radiation damage.
Nonuniformly irradiated sensors allowed the study of charge collection and \ttt variations as a function of fluence.
In particular there is a clear enhancement of charge collected from depths up to $80 \mum$ at an operating voltage of 1000 V due to charge multiplication for fluences higher than $6\times \fluence$.
The charge multiplication effect has been observed for proton but not neutron irradiated sensors.

%%%%%%%%%%%%%%%%%%%%%%%%%%%%%%%%%%%%%%%%%%%%%%%%%%%%%%%%%%%%%%%%%%%%%%%%%%%%%%%%%%%%%%%%
\section*{Acknowledgements}

We would like to express our gratitude to our colleagues in the CERN accelerator departments for the excellent performance of the beam in the SPS North Area.
We would like to acknowledge  Eugenia Price for her work on the slow controls of the telescope and  DUTs, Daniel Saunders for his contributions to the online data monitoring and tracking,  Mark Williams for his work on the test pulse calibration of the assemblies,  and Jan Buytaert, Wiktor Byczynski and Raphael Dumps for their extensive and continuous support to keep the telescope operational.  We would also like to thank all people that took part in the 
data taking effort throughout the years of 2014 to 2016.
We gratefully acknowledge the financial support from CERN and from the national agencies: CAPES, CNPq, FAPERJ (Brazil); the Netherlands Organisation for Scientific Research (NWO); The Royal Society and the Science and Technology Facilities Council (U.K.). 
This project has received funding from the European Union’s Horizon 2020 Research and Innovation programme under Grant Agreement no. 654168 and from the People Programme (Marie Curie Actions) of the European Union’s Seventh Framework Programme FP7/2007-2013/ under REA grant agreement nr. 317446 INFIERI “INtelligent Fast Interconnected and Efficient Devices for Frontier Exploitation in Research and Industry”.

\clearpage

\appendix

\section{List of assemblies}
\label{app:assemblies}

The details of the assemblies tested are summarised in \tab\ref{tab:app_assemblies}. 
\begin{table}[h!]
\caption{Assemblies tested.}
\label{tab:app_assemblies}
\centering
\scalebox{0.9}{
\begin{tabular}{lccccccc}
  \toprule
  ID   & Vendor & \makecell{Thickness \\ {[$\!\mum$]} } & Type  & \makecell{Edge width \\  {[$\!\mum$]} } & \makecell{Implant \\  {[$\!\mum$]} } & \makecell{Irradiation \\ facility } & \makecell{Peak fluence \\  {[\fluence]} } \\
  \midrule
  S6  & HPK   & 200 & \np & 450 & 39 & JSI & 8 \\
  S8  & HPK   & 200 & \np & 450 & 35 & IRRAD & 8 \\
  S11  & HPK   & 200 & \np & 450 & 39 & IRRAD & 8 \\
%  \textbf{S15}  & HPK   & 200 & \np & 450 & 35 & JSI & 4 \\
  S17  & HPK   & 200 & \np & 450 & 39 & JSI & 8 \\
%  \textbf{S18}  & HPK   & 200 & \np & 450 & 39 & - & - \\
%  \textbf{S20}  & HPK   & 200 & \np & 450 & 35 & - & - \\
  S22  & HPK   & 200 & \np & 450 & 35 & JSI & 8 \\
  S23  & Micron   & 200 & \np & 450 & 36 & JSI & 8 \\
  S24  & Micron   & 200 & \np & 450 & 36 & JSI & 8 \\
  S25  & Micron   & 200 & \np & 450 & 36 & IRRAD & 8 \\
  S27  & Micron   & 150 & \nn & 450 & 36 & JSI & 8 \\
  S29  & Micron   & 150 & \nn & 450 & 36 & JSI & 8 \\
  S30  & Micron   & 150 & \nn & 450 & 36 & IRRAD & 8 \\
 % \textbf{S31}  & Micron   & 200 & \np & 250 & 36 & - & - \\
  S33  & Micron   & 150 & \nn & 250 & 36 & - & - \\
  S34  & Micron   & 150 & \nn & 250 & 36 & - & - \\
  \bottomrule
  \end{tabular}
  }
\end{table}

%%%%%%%%%%%%%%%%%%%%%%%%%%%%%%%%%%%%%%%%%%%%%%%%%%%%%%%%%
\section{Timewalk correction}
\label{app:timewalk}

\begin{figure}[h!]
  \centering
  {\includegraphics[width=0.49\textwidth]{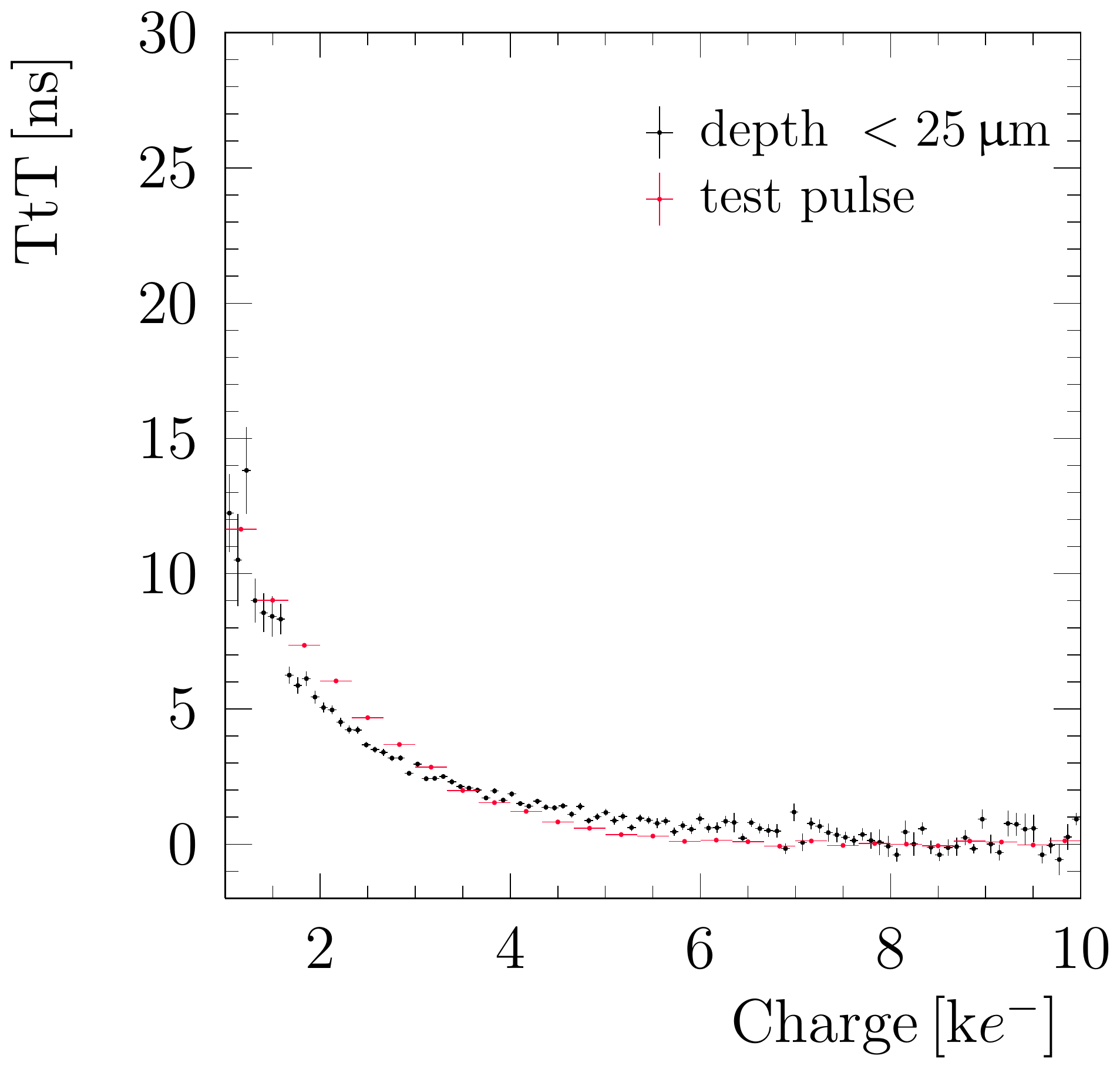}}
  \caption{Comparison between timewalk curve from test pulse data and from test beam data with charges liberated close to the electrode. 
  The sensor is a $200 \mum$ thick nonuniformly irradiated HPK \np sensor at 1000~V bias voltage. }
  \label{fig:tw_tp_comparison}
\end{figure}

With the grazing angle method, the timewalk curve can be determined directly from the testbeam data by selecting only charges liberated at small depth, up to about $25 \mum$ from the pixel electrodes. 
The timewalk curve obtained is validated for some assemblies by comparing it to the timewalk curve determined by injecting a test pulse with known charge in the pixel front-end, as shown in \fig\ref{fig:tw_tp_comparison}. 
The shape variation of the timewalk curve is negligible, leading to the conclusion that the profile obtained is sufficiently representative of a pure electronics effect. The horizontal bar on the hit charge of the test pulse curve is due to the binning and it is not representative of the charge uncertainty.

\begin{figure}[h!]
  \centering
  {\includegraphics[width=0.49\textwidth]{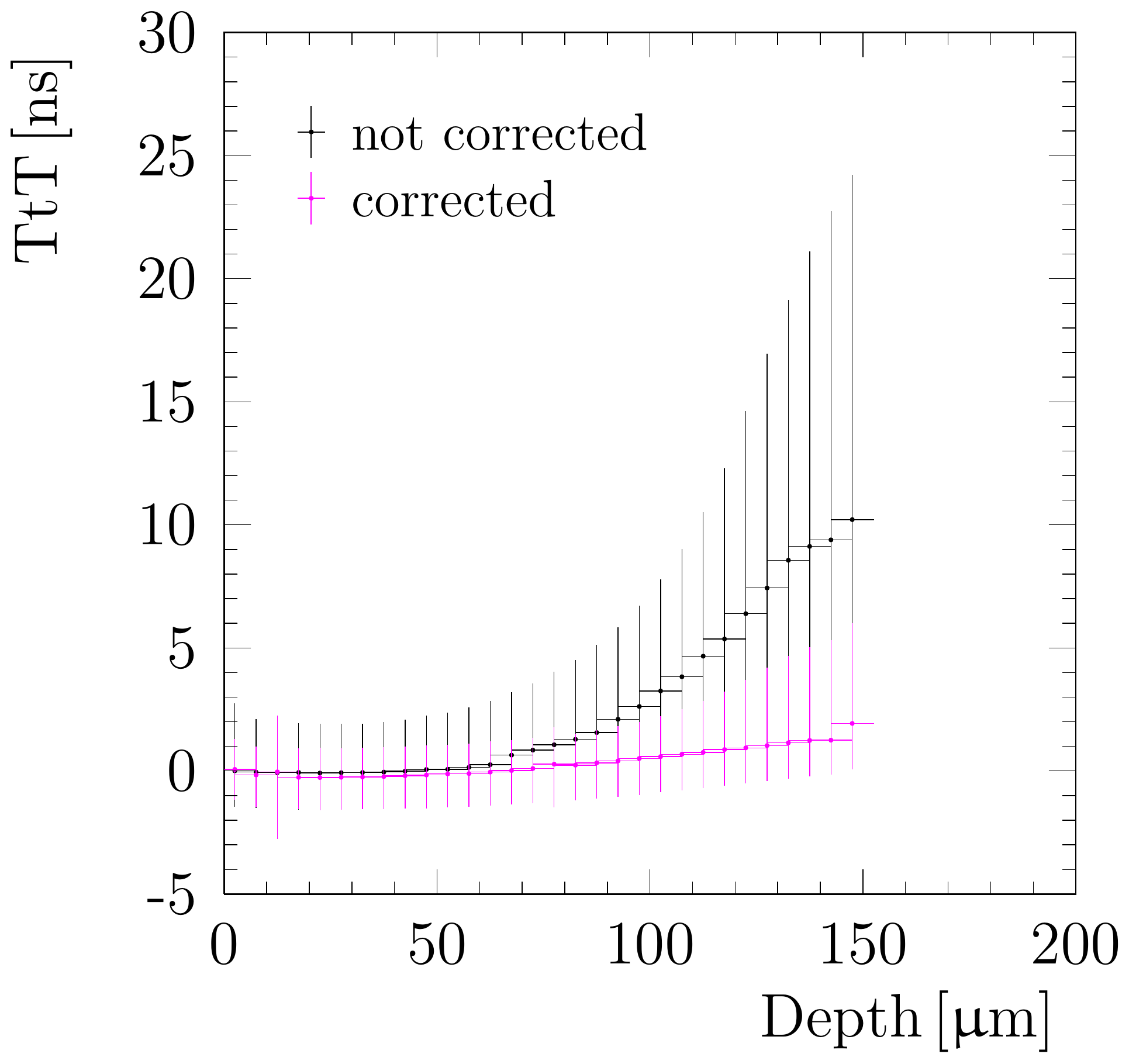}} \hspace{1mm}
  {\includegraphics[width=0.49\textwidth]{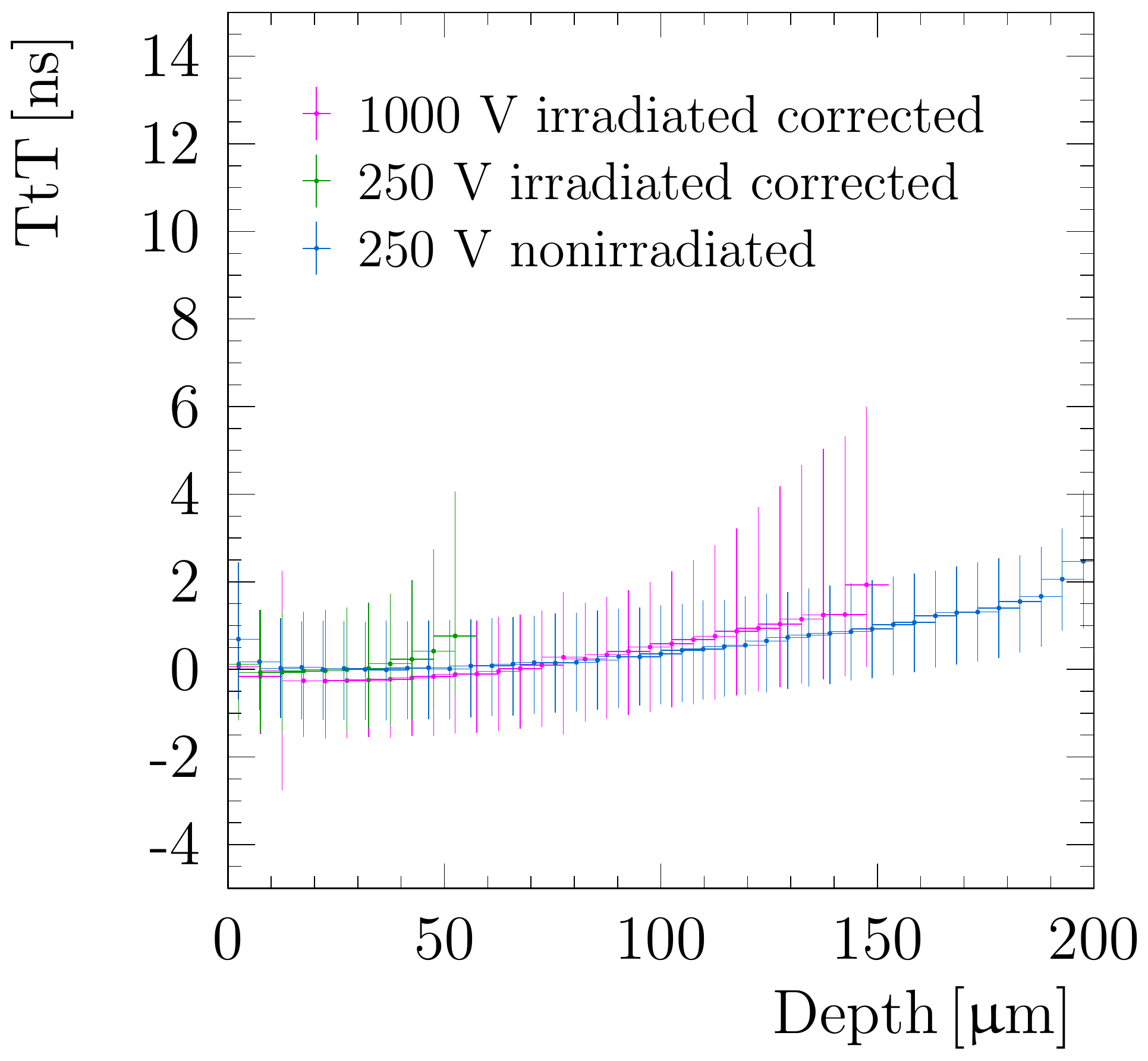}} 
  \caption{ Comparison between \ttt profiles before and after timewalk correction (left) and  comparison between \ttt profiles for a nonirradiated sensor operated at 250 V and for a uniformly irradiated sensor operated at 250 V and 1000 V (right). The sensors are $200 \mum$ thick HPK \np sensors. The uncertainty is assigned as uncertainty on the single measurement.}
  \label{fig:timewalk_corr}
\end{figure}

For assemblies irradiated at full fluence, the contribution of timewalk becomes significant, since the charge collected is lower than 3000~\en.
The timewalk effect broadens the \ttt distribution, leading to an asymmetric uncertainty on the single measurement that varies from 3~\ns up to 15~\ns depending on the depth.
In the left plot of \fig\ref{fig:timewalk_corr} the \ttt profile for a HPK \np sensor at 1000~V is compared to the profile obtained by applying the timewalk correction described in \sect\ref{sec:method}.
The timewalk curve is fitted and a correction to the \ttt of each hit is applied as a function of charge.
The magnitude of the correction increases as the charge decreases, hence with depth, leading to a smaller mean value and narrower distribution compared to the uncorrected case. 
The corrected \ttt spread results in less than 3~\ns along the whole sensor depth and the uncertainty spans from $\sim 1.5 \ns$ at small depth up to $\sim 4 \ns$ at the border of the active region.
In the right plot of \fig\ref{fig:timewalk_corr} the corrected profiles at the highest (1000~V) and lowest (250~V) voltage tested are compared to the \ttt profile for the same type of sensor before irradiation and fully depleted.
The profiles agree within the uncertainties. 
The main difference with respect to the nonirradiated case is represented by the larger uncertainty on the single measurement: close to the electrodes the uncertainty is comparable, while at the border of the active region is almost double, due to possible residual timewalk.


\section{Rise in MPV close to threshold}
\label{app:threshold}
\begin{figure}[b]
    \centering
    \includegraphics[width=0.49\textwidth]{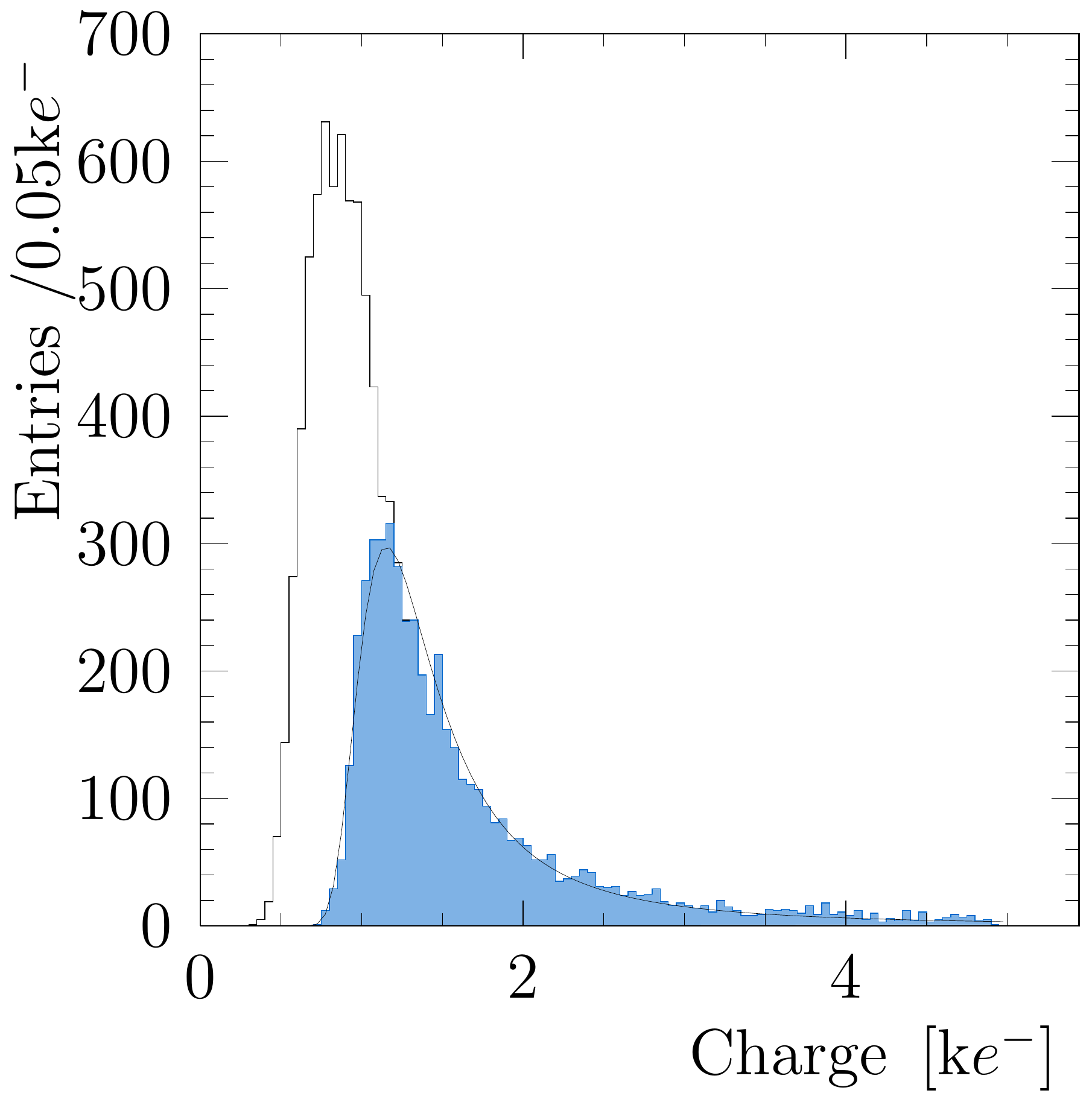} \hspace{1mm}
     \includegraphics[width=0.49\textwidth]{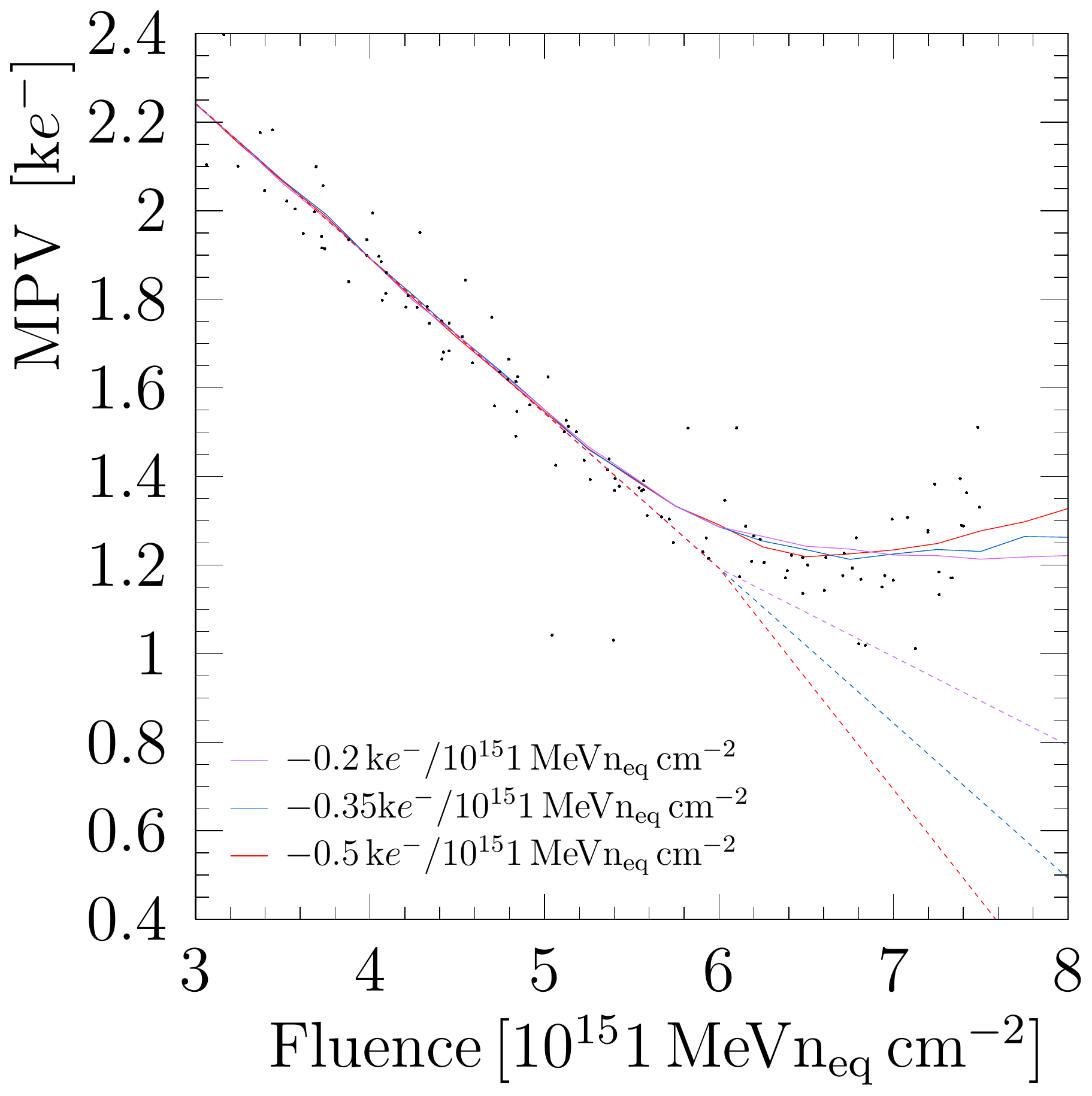} 
    \caption{Left: Distribution of simulated of charges for an MPV of $800e^{-}$, demonstrating the effect of the threshold on the shape, with the filled histogram indicating the charges that pass the threshold. Right: Comparison with trend observed in a 200$\mum$ HPK n-on-p sensor operated at 750~V at a depth of 114$\mum$, assuming different linear dependence on the true MPV with fluence. }
    \label{fig:app3_fig1}
\end{figure}

A small rise in the measured MPV as a function of fluence is observed in the grazing angle setup close to the threshold at some depths, as shown in Fig.\ref{fig:MPVvsFluenceS8}. 
The MPV in some cases should be below the threshold from a naive extrapolation from lower fluences, and thus the expected charge distribution in such cases is investigated using pseudoexperiments. 
The distribution of charges is generated using parameters typical of the real data, and the threshold emulated by selecting only charges above ${1000\pm100}e^{-}$, where the variation is assumed to be gaussian in nature. 
Both the true charge distribution and that with the emulated threshold are shown in Fig.~\ref{fig:app3_fig1} for an MPV of $800\electron^{-}$. The charge distribution can still be described by a Landau function after the threshold has been applied, albeit with a significantly larger MPV and width. Due to the wide tail of the Landau distribution, the measured MPV increases as the true MPV decreases, up to a few times the threshold dispersion. 

This model is compared to the data set where the effect appears most pronounced, at a depth of $114\mum$ and an applied voltage of 750~V. The true MPV is estimated by assuming a linear decrease as a function of fluence, with slopes of: ${-0.2,-0.35,-0.5 \text{k}e^{-}/{10^{15} 1 \mev \text{n}_{\text{eq}}\cm^{-2}}}$ considered. The MPV obtained by performing a fit to the simulated data sets are compared with real data in Fig.~\ref{fig:app3_fig1}, where the observed rise in MPV is largest when the true MPV is furthest below threshold. There is qualitatively a good agreement between the data and the model, and thus this effect can account for the artificial rise in MPV at very high fluences and close to the threshold.

\clearpage
\addcontentsline{toc}{section}{References}
\setboolean{inbibliography}{true}
\bibliographystyle{JHEP}
\bibliography{main,LHCb-PAPER,LHCb-CONF,LHCb-DP,LHCb-TDR}
% \bibliography{LHCb-PAPER}
\end{document}